\documentclass[12pt, draftclsnofoot, onecolumn]{IEEEtran}

\makeatletter
\def\ps@headings{%
\def\@oddhead{\mbox{}\scriptsize\rightmark \hfil \thepage}%
\def\@evenhead{\scriptsize\thepage \hfil \leftmark\mbox{}}%
\def\@oddfoot{}%
\def\@evenfoot{}}
\makeatother 

\usepackage{graphicx}
\usepackage{amsmath, amsfonts,epsfig, multirow, floatflt}
\usepackage{subfigure}
\usepackage{cite}
\usepackage{booktabs}
\usepackage{verbatim}
\usepackage{amsthm}
\usepackage{color}

\begin{document}

\title{Optimal Reliability in Energy Harvesting Industrial Wireless Sensor Networks}

\author{Lei~Lei {\it Member, IEEE}, Yiru~Kuang, Xuemin (Sherman) Shen$^\ast$ {\it Fellow, IEEE}, Kan Yang$^\ast$ {\it Member, IEEE}, Jian~Qiao$^\ast$, and Zhangdui~Zhong

\small State Key Laboratory of Rail Traffic Control \& Safety\\
Beijing Jiaotong University, Beijing, China, 100044\\
leil@bjtu.edu.cn

$^\ast$Department of Electrical and Computer Engineering\\
University of Waterloo, Waterloo, Ontario, Canada~~N2L 3G1
}

%
%
%
%
%
%

\maketitle

\begin{abstract}
For Industrial Wireless Sensor Networks, it
is essential to reliably sense and deliver the environmental data
on time to avoid system malfunction.
While energy harvesting is a promising technique to extend
the lifetime of sensor nodes, it also
brings new challenges for system reliability due to the stochastic
nature of the harvested energy. In this paper, we investigate
the optimal energy management policy to minimize
the weighted packet loss rate under delay constraint, where the packet
loss rate considers the lost packets both during the sensing and delivering
processes. We show that the above energy management problem can be modeled as an infinite horizon average reward constraint Markov decision problem. In order
to address the well-known curse of dimensionality problem and
facilitate distributed implementation, we utilize the linear value approximation
technique. Moreover, we apply stochastic online learning with
post-decision state to deal with the lack of knowledge of the
underlying stochastic processes. A distributed energy allocation algorithm with water-filling structure and a scheduling algorithm by auction mechanism are obtained. Experimental results show that
the proposed algorithm achieves nearly the same performance
as the optimal offline value iteration algorithm while requiring
much less computation complexity and signaling overhead, and
outperforms various existing baseline algorithms.
\end{abstract}


\section{Introduction}
Industrial wireless sensor networks (IWSNs) are the integration of wireless sensor networks (WSNs) and industrial systems,
in which wireless sensor nodes are installed on industrial equipments to monitor their conditions or efficiency through various
parameters, such as vibration, temperature, pressure, and power quality \cite{CST:Kumar}. These data are then transmitted over the air to a sink node, which is connected to a control system, for further analysis. Based on the analysis result from the sink node, the control system can control actuators in a machine or alert users. Due to the collaborative intelligence and low-cost nature of IWSNs, it is widely adopted in industrial applications to achieve flexibility, self-configuration, rapid deployment, intelligent controlling, and an inherent intelligent processing capability.\par

Recently, energy harvesting (EH) has emerged as a promising technique to extend the lifetime of sensor nodes with rechargeable batteries
by harvesting the available ambient energy (e.g., solar, motion, heat, aeolian etc.), especially when battery replacement is difficult
or cost-prohibitive \cite{Sudevalayam:CST}. In energy harvesting industrial wireless sensor networks (EH-IWSNs), energy conservation is no longer the prime design issue since the sensor nodes can theoretically operate over an unlimited time horizon with the renewable energy. However, achieving high reliability in EH-IWSNs is a challenging technical issue due to the uncontrollable and unstable nature of the harvested energy arrival. Therefore, the energy management strategy for an EH-IWSN needs to take into account the energy replenishment process, so that the long-term reliability performance of the overall system in regard to sensing and data communication tasks can be maximized by taking full advantage of the EH process and simultaneously avoid premature energy depletion before the next recharge cycle.\par

In general, the reliability of EH-IWSN systems is essential for many industrial applications, which means that data received at the sink node (the control center) must accurately reflect what is actually occurring in the industrial environment. The reliability of EH-IWSN systems depends on both the sensing process and transmission process, which means that the environmental data should be \emph{reliably} captured by the sensor nodes and the sensed data should be \emph{reliably} transmitted to the sink. Energy management is a promising approach to deal with this technical challenge. Consider an EH-IWSN with a finite data buffer and a finite battery energy buffer for each energy harvesting sensor (EHS) node. If the EHS node reserves an excessive amount of energy for sensing and leaves an insufficient amount of energy for transmission, the newly sensed data may be dropped by the data buffer due to its limited capacity. On the other hand, if an excessive amount of energy is consumed for transmission, there may not be enough sensing power left to capture the environmental data. In addition, if the energy allocation at the current decision epoch is overly aggressive, the EHS node may stop functioning at the next decision epoch because of the energy outage. Besides, since sensor data is typically time-sensitive, e.g., alarm notifications for the industrial facilities, it is also important to receive the data at the sink in a timely manner. Delayed, incorrectly received or lost data may cause industrial applications to malfunction, and lead to wrong decisions in the monitoring system \cite{TII:Heo}.\par

In this paper, we consider an IWSN where a fusion center (FC) collects data from multiple EHS nodes with slot-by-slot transmission. Each sensor node has an energy buffer and a data buffer with finite size. A random number of packets from the industrial environment should be sensed at each EHS node during each time slot. Moreover, the scheduled sensor node along with the allocated energy for data transmission need to be determined by taking into account the battery energy state information (BSI), queue state information (QSI), and channel state information (CSI) at the beginning of the time slot. The remaining energy in the battery can be used to sense the packets throughout the time slot. Ideally, the EHS nodes should sense all these packets and transmit them to the FC without any loss or error within the delay constraint. However, packets may be lost during data sensing (due to limited sensing power), and data communication (due to both the effect of queuing overflow in the Medium Access Control (MAC) layer, and the packet reception error in the physical layer). Our objective is to minimize the weighted packet loss rate in the system under per-node delay constraints, where the weight of the packet loss rate of every EHS node is used to model the different reliability requirements of different sensors. In other words, we aim at maximizing the system reliability while guaranteeing real-time transmission.\par

Specifically, we formulate the reliability optimal energy management problem in EH-IWSNs and solve it by casting it into an infinite horizon average reward constrained Markov Decision Process (CMDP). The main contributions of this paper lie in the following aspects.\par

\begin{enumerate}
  \item \emph{Reliability modeling}: Packet loss rate is widely used in WSNs to quantify system reliability, where most previous work focuses on reliable data transmission under wireless channel errors \cite{SECON:Kim}. \emph{Different from these works, the main contribution of our reliability model lies in that the reliability of data sensing and data communication subsystems are jointly considered}.
   \item \emph{Low complexity distributed control with local system state}: Markov Decision Process (MDP) is a systematic approach in dealing with the dynamic optimization problem, to which the resource control problem in EH-WSNs belongs due to the dynamic energy arrival and time-varying wireless channel. However, although the MDP method is widely adopted in point-to-point wireless communication scenario \cite{Castagnetti:TII,TWC:Sharma,TWC:Blasco,TN:Srivastava,TVT:Mao}, it is generally not used in existing literature considering multiple nodes in EH-WSN due to the curse of dimensionality that forbids its practical implementation \cite{Gul:WCNC,Sharma:Allerton,Joseph:MSWiM,Testa:ISWCS,INFOCOM:Liu,TN:Huang,TWC:Gatzianas,Cai:JSAC}. \emph{The first contribution related to the solution of the MDP model lies in that in order to deal with the curse of dimensionality problem, we derive an equivalent Bellman's equation with reduced state space and approximate the global value functions by a sum of per-node value functions, which can be distributively maintained at every EHS node.} Therefore, our proposed energy allocation action can be locally determined by an EHS node based on its local state observation using a simple formula with multilevel water-filling structure. Moreover, the scheduling action is determined by an auction mechanism, where each EHS node computes and sends its bid to the FC. In this way, the signaling overhead is greatly reduced compared to a centralized solution, where the sensor nodes have to send their QSI and BSI to the FC.
   \item \emph{No information requirement about the underlying stochastic processes}: Due to the harsh radio propagation environment in IWSNs and the unpredictable nature of the EH and packet arrival rate, an explicit knowledge of the probability distributions of the underlying stochastic processes may not be available to the FC and EHS nodes \cite{TWC:Gatzianas}. Moreover, information about the packet/energy arrival amounts are only available at the end of every time slot since they are being generated throughout the period. \emph{The second contribution related to the solution of the MDP model lies in that we utilize post-decision state and stochastic online learning framework so that our proposed algorithm does not require the above information of the underlying stochastic processes.}
\end{enumerate}

The remainder of the paper is organized as follows. The related works are reviewed in Section II. In Section III, we introduce our system model. In Section IV, we elaborate the MDP problem formulation and derive the low-complexity near-optimal control policy in Section V. We discuss the performance simulations in Section VI. Finally, we summarize the main results and discuss our future work in Section VII.

\section{Related Work}
In recent years, WSNs with EHS nodes have attracted a lot of attention. There are intense research efforts in the design of optimal energy management policies for a single EHS node, e.g., \cite{Castagnetti:TII,TWC:Sharma,JSAC:Ozel,TWC:Blasco,TN:Srivastava,TVT:Mao}. In \cite{Castagnetti:TII}, a joint duty-cycle optimization and transmission power control approach is proposed to maximize the number of transmitted packets while respecting the limited and time-varying amount of available energy. In \cite{TWC:Sharma}, the throughput optimal and delay optimal energy management policies are derived assuming the data buffer and energy storage buffer are both infinite. In \cite{JSAC:Ozel}, online transmission policies with the objective of maximizing the deadline constrained throughput under channel fluctuations and energy variations are studied assuming a finite energy queue but an infinite data queue. In \cite{TWC:Blasco}, a learning theoretic optimization approach to maximize the expected total transmitted data during the transmitter's activation time is studied, which assumes that the rechargeable battery has a finite-capacity, and a data packet arrives at the beginning of a time slot is lost if not transmitted within the following time slot. \cite{TN:Srivastava} addresses the problem of energy management for energy-replenishing nodes with finite battery and finite data buffer capacities, and gives an energy management scheme that achieves the optimal utility asymptotically while keeping both the battery discharge and packet loss probabilities low. \cite{TVT:Mao} studies the energy allocation problem for sensing and transmission in an energy harvesting sensor node, which takes into account both the data sensing energy consumption and the finite capacity of the data buffer.\par


Compared with the vast literature on single node scenario, the problem of analyzing and modeling the interaction among multiple EH nodes at the MAC layer in a network has received limited attention so far. \cite{Gul:WCNC} considers a single-hop TDMA wireless network where a FC collects data from multiple EHS nodes with infinite battery size, and find a low complexity scheduling policy that maximizes the total throughput of the data backlogged system under the assumption that unit energy is consumed for the transmission of one packet. The authors in \cite{Sharma:Allerton} discuss the extension of their proposed energy management policies to multiple EHS nodes scenario. The joint power control, routing and scheduling protocols are proposed in \cite{Joseph:MSWiM} for multihop sensor networks. In \cite{Testa:ISWCS}, the problem of optimizing the transmission strategy of the two nodes over a shared wireless channel by a central controller is considered, with the goal of maximizing the long-term average importance of the transmitted data. In \cite{INFOCOM:Liu}, distributed routing, rate control and scheduling algorithm for energy-harvesting sensor networks is considered using the dual decomposition and subgradient method. All the above research \cite{TWC:Sharma,Joseph:MSWiM,Gul:WCNC,Testa:ISWCS,INFOCOM:Liu} considers there are infinite backlogs of packets at the transmitter and does not consider the delay constraint of the transmitted packet. On the other hand, \cite{TN:Huang,TWC:Gatzianas} consider the dynamic data arrival and use the Lyapunov drift and optimization framework to develop throughput-optimal (in stability sense) resource
allocation algorithm for wireless networks operating with rechargeable batteries. However, the data buffer is assumed to be infinite in \cite{TN:Huang,TWC:Gatzianas}, and real-time transmission requirement and sensing power are not considered. When dealing with the multiple EHS nodes scenario, MDP is generally not used due to its curse of dimensionality problem. In \cite{Cui:TSP}, a delay-optimal base station discontinuous transmission (BS-DTX) control and user scheduling for downlink coordinated multiple-input and multiple-output (MIMO) systems with energy harvesting capability is studied using MDP method.\par


The above approaches cannot be directly applied to the IWSNs, since they do not consider the reliability and real-time requirements which are vital for many industrial applications. In \cite{TII:Heo}, a novel routing protocol
for IWSNs which provides real-time, reliable delivery of a packet, while considering energy awareness is proposed. In \cite{Luo:TII}, an opportunistic routing algorithm to minimize energy consumption and maximize network lifetime of WSN is proposed. However, they only consider non-rechargeable batteries and without considering the MAC layer mechanisms.\par


\section{System Model}
\begin{figure}
\centering
\includegraphics[width=0.6\textwidth]{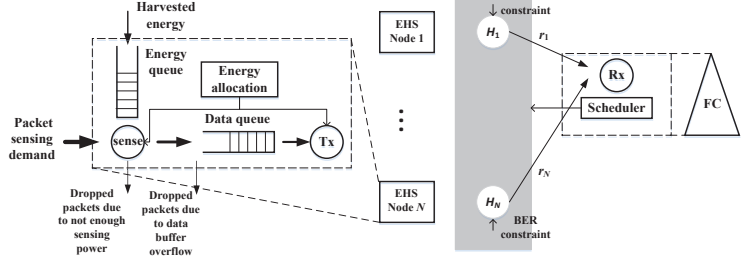}
\caption{A single-hop IWSN where a FC collects data from $N$ EHS nodes.}
\label{systemmodel}
\end{figure}

We consider a single-hop IWSN where a FC collects data from $N$ EHS nodes, as illustrated in Fig.\ref{systemmodel}. The EHS nodes perform sensing tasks and generate packets to be transmitted to the FC. The IWSN operates in a TDMA fashion over time slots of equal duration $\tau$. In each time slot, the FC schedules one EHS node for data transmission. The EHS node has a rechargeable battery with capacity $B_{\mathrm{max}}$ energy units and a data buffer with size $Q_{\mathrm{max}}$ packets, where all packets have fixed length of $K$ bits. Each EHS node collects energy from the environment, which is stored in the rechargeable battery for sensing and data transmission. In addition, each EHS node performs sensing in the field, stores the sensed data in the data buffer, and transmits the data to the FC over a wireless channel when scheduled.\par

We consider an additive white Gaussian noise (AWGN) channel with flat fading, which is modeled as a first-order Finite State Markov Chain (FSMC) \cite{SPM:Sadeghi,TWC:Kan}. Define the CSI of EHS node $n\in\{1,\ldots,N\}$ to be $H_{n,t}\in\mathcal{H}$, $t=0,1,2,\ldots$, which takes a value from the discrete state space $\mathcal{H}$ and denotes the channel gain of EHS node $n$ at time slot $t$. The CSI remains constant within a time slot and the CSI at time slot $t$ only depends on the CSI at time slot $t-1$. We assume that the EHS node $n$ knows its local CSI $H_{n,t}$ at the beginning of time slot $t$. Due to its high spectral efficiency, we consider the multi-level quadrature amplitude modulation ($M$-QAM) is used for adaptive modulation. Consider the scheduled EHS node at time slot $t$, if the allocated transmission energy is $p_{n,t}^{\mathrm{T}}$, then the average instantaneous transmission power is $p_{n,t}^{\mathrm{T}}/\tau$. Given a target bit error rate (BER) $\epsilon_{n}$, the EHS node is able to transmit $r(H_{n,t},p_{n,t}^{\mathrm{T}})$ packets during time slot $t$ as approximated by
\begin{equation}
\label{eq1}
r(H_{n,t},p_{n,t}^{\mathrm{T}})=\frac{\tau W}{K}\log_{2}(1+\xi_{n}\frac{H_{n,t} p_{n,t}^{\mathrm{T}}}{N_{0}W\tau}),
\end{equation}
\noindent where $N_{0}$ is the power spectral density of the Gaussian noise and $W$ is the bandwidth of the channel. $\xi_{n}=-\frac{c_{2}}{\ln(\epsilon_{n}/c_{1})}$ with $c_{1}\approx 0.2$ and $c_{2}\approx 1.5$ for small BER \cite{JSAC:Han}. \par

Define the QSI of EHS node $n\in\{1,\ldots,N\}$ to be $Q_{n,t}$, which denotes the number of packets stored at the data buffer of EHS node $n$ at the beginning of the time slot $t$. Define the BSI of EHS node $n\in\{1,\ldots,N\}$ to be $B_{n,t}$, where $B_{n,t}$ denotes the number of harvested energy units of EHS node $n$ at the beginning of time slot $t$. Denote the number of packets that arrive at EHS node $n$ (a packet ``arrives at EHS node $n$" means that the packet needs to be sensed by EHS node $n$, but have not been sensed yet) during time slot $t$ as $A_{n,t}$. We consider $A_{n,t}$ is i.i.d. over time slots and independent w.r.t. $n$ according to a general distribution $f_{A}$ with average arrival rate $\mathbf{E}[A_{n}]=\lambda_{n}^{\mathrm{A}}$, and the statistical characteristics of $A_{n,t}$ is unknown to the EHS node $n$. Moreover, we consider the units of energy needed to sense $A_{n}$ packets is $A_{n}/\gamma$\footnote{In general, the amount of data generated $\nu(p)$ is a monotonically non-decreasing and concave function in the units of energy $p$ used for sensing. In this paper, we consider that $\nu(p)$ is a linear function of $p$, although our method is applicable to other forms of $\nu(p)$ as well.}, where $\gamma$ is the data sensing efficiency parameter (i.e., the number of packets that the sensor can sense per unit energy)\cite{INFOCOM:Chen}. Since at the beginning of time slot $t$, $p_{n,t}^{T}$ units of energy is allocated for data transmission, the available units of energy left for sensing during the whole time slot $t$ is $B_{n}-p_{n,t}^{T}$, and the actual number of packets that are sensed during time slot $t$ is $\min[A_{n,t},\gamma(B_{n}-p_{n,t}^{T})]$, which can only be observed by the EHS node $n$ at the end of time slot $t$. The packets obtained by sensing at time slot $t$ will be queued up in the data buffer until they are transmitted in the subsequent time slots. If the queue length reached the buffer size $Q_{\mathrm{max}}$, the subsequent sensed packets will be dropped. According to the above assumption, the queuing process evolves as follows:
\begin{align}
\label{eq2}
Q_{n,t+1}=\min[Q_{\mathrm{max}},\max[0,Q_{n,t}-r(H_{n,t},p_{n,t}^{\mathrm{T}})]+\min[A_{n,t},\gamma(B_{n}-p_{n,t}^{T})]].
\end{align}

Define the harvested energy arrival of EHS node $n\in\{1,\ldots,N\}$ to be $\{E_{n,t}\}_{t=0,1,\ldots}$, which is considered to be i.i.d over scheduling slots and independent w.r.t. EHS node $n$ according to a general distribution $f_{E}$ with average arrival rate $\mathbf{E}[E_{n}]=\lambda_{n}^{\mathrm{E}}$, and the statistical characteristics of $E_{n,t}$ is unknown to the EHS node $n$. During the whole time slot $t$, the EHS node $n$ is able to recharge energy by $E_{n,t}$, which can be used for sensing or transmission in time slot $t+1$ onward. As a result, the EHS node does not know the value of $E_{n,t}$ until the end of time slot $t$. During time slot $t$, on the one hand, the EHS node harvests $E_{n,t}$ units of energy from the environment. On the other hand, it consumes $p_{n,t}^{\mathrm{T}}$ units of energy for data transmission and $\min[A_{n,t}/\gamma,B_{n,t}-p_{n,t}^{\mathrm{T}}]$ units of energy for sensing. Since the rechargeable battery has a finite capacity $B_{\mathrm{max}}$, the energy stored in the battery is updated as follows:
\begin{equation}
\label{eq3}
B_{n,t+1}=\min\left[B_{\mathrm{max}},\max[0,B_{n,t}-p_{n,t}^{\mathrm{T}}-A_{n,t}/\gamma]+E_{n,t}\right].
\end{equation}

\section{Problem Formulation}
In this section, we shall formulate the problem of minimizing the weighted packet loss rate under the delay constraint using infinite-horizon average reward CMDP model, which consists of four elements: states, actions, state transition probabilities, and rewards.\par

The global system state at time slot $t$ can be characterized by the aggregation of the system CSI, QSI and BSI, which are denoted as $\mathbf{S}_{t}=(\mathbf{H}_{t},\mathbf{Q}_{t},\mathbf{B}_{t})$, where $\mathbf{H}_{t}=\{H_{n,t}\}_{n=1}^{N}$, $\mathbf{Q}_{t}=\{Q_{n,t}\}_{n=1}^{N}$, and $\mathbf{B}_{t}=\{B_{n,t}\}_{n=1}^{N}$. $\mathbf{H}_{t}$, $\mathbf{Q}_{t}$ and $\mathbf{B}_{t}$ take discrete values and are all bounded. Let $\mathcal{S}=\mathcal{H}\times\mathcal{Q}\times\mathcal{B}$ be the full system state space, which is discrete and countable. \par

At each time slot $t$, based on the current state $\mathbf{S}_{t}$, an action $\mathbf{a}_{t}=\{\mathbf{x}_{t},\mathbf{p}_{t}^{\mathrm{T}}\}$ is taken from the set of allowable actions in the action space $\mathcal{A}$, which is discrete and finite. The action is composed of scheduling action $\mathbf{x}_{t}:=\{x_{n,t}\in \{0,1\}| \sum_{n=1}^{N}x_{n,t}\leq1\}_{n=1}^{N}$, as well as transmission energy allocation action $\mathbf{p}_{t}^{\mathrm{T}}:=\{p_{n,t}^{\mathrm{T}}|0\leq p_{n,t}^{\mathrm{T}}\leq B_{n,t}\}_{n=1}^{N}$. Note that the transmission energy allocation and scheduling action are correlated, since $p_{n,t}^{\mathrm{T}}=0$ if $x_{n,t}=0$ and $p_{n,t}^{\mathrm{T}}>0$ if $x_{n,t}=1$, i.e., $x_{n,t}=\mathbf{I}\{p_{n,t}^{\mathrm{T}}>0\}, \ \forall n=1,\ldots,N$. 
A control policy prescribes a procedure for action selection in each state at all decision epoches $t$. To facilitate implementation, we consider stationary Markovian deterministic control policies. A deterministic Markovian control policy given by $\Omega$ is a mapping $\mathcal{S}\rightarrow\mathcal{A}$ from the state space to the action space, which is given by $\Omega(\mathbf{S})=\mathbf{a}\in\mathcal{A}, \ \forall \mathbf{S}\in\mathcal{S}$. Such a policy is said to be Markovian because it depends on previous systems and actions only through the current state of the system. Stationary policies are fundamental to the theory of infinite horizon Markov decision processes, in which the mapping function $\Omega$ does not change with time.\par

The induced random process can be represented by the discrete-time Markov chain (DTMC) $\{\mathbf{S}_{t}\}_{t=0,1,\ldots}$. Given a system state $\mathbf{S}_{t}$ and an action $\mathbf{a}_{t}$ at time slot $t$, the state transition probability of the DTMC is given by
\begin{align}
\label{eq4}
\mathrm{Pr.}\{\mathbf{S}_{t+1}|\mathbf{S}_{t},\mathbf{a}_{t}\}=\mathrm{Pr.}\{\mathbf{H}_{t+1}|\mathbf{H}_{t}\}\mathrm{Pr.}\{\mathbf{Q}_{t+1}|\mathbf{H}_{t},\mathbf{Q}_{t},\mathbf{B}_{t},\mathbf{a}_{t}\}\mathrm{Pr.}\{\mathbf{B}_{t+1}|\mathbf{B}_{t},\mathbf{a}_{t}\},
\end{align}
\noindent where $\mathrm{Pr.}\{\mathbf{Q}_{t+1}|\mathbf{H}_{t},\mathbf{Q}_{t},\mathbf{B}_{t},\mathbf{a}_{t}\}$ and $\mathrm{Pr.}\{\mathbf{B}_{t+1}|\mathbf{B}_{t},\mathbf{a}_{t}\}$ can be derived from \eqref{eq2} and \eqref{eq3}, respectively.\par

Given a deterministic control policy $\Omega$, since the action $\mathbf{a}_{t}$ under every system state $\mathbf{S}_{t}$ is determined, we can directly derive $\mathrm{Pr.}\{\mathbf{S}_{t+1}|\mathbf{S}_{t},\Omega(\mathbf{S}_{t})\}$.
Define the transition probability matrix $\mathbf{P}^{\Omega}=[\mathrm{Pr.}\{\mathbf{S}_{t+1}=\mathbf{S}'|\mathbf{S}_{t}=\mathbf{S},\Omega(\mathbf{S})\}],\mathbf{S},\mathbf{S}'\in\mathcal{S}$ and the steady-state probability matrix $\boldsymbol{\pi}^{\Omega}=[\pi_{\mathbf{S}}^{\Omega}],\mathbf{S}\in\mathcal{S}$, where $\pi_{\mathbf{S}}^{\Omega}=\lim_{t\rightarrow\infty}\mathrm{Pr.}\{\mathbf{S}_{t}=\mathbf{S}\}$. Each element of the transition probability matrix $\mathbf{P}^{\Omega}$ can be derived. Then, the stationary distribution of the ergodic process $\{\mathbf{S}_{t}\}_{t=0,1,\ldots}$ can be uniquely determined from the balance equations when the Markov chain is irreducible. As we deal with the finite- and countable-state model in this paper, all stationary policies generate Markov chains with a single irreducible class \cite{book:Puterman}. Denote the global CSI, QSI and BSI under system state $\mathbf{S}$ (resp. $\mathbf{S}'$) by $\mathbf{H}$ (resp. $\mathbf{H}'$), $\mathbf{Q}$ (resp. $\mathbf{Q}'$) and $\mathbf{B}$ (resp. $\mathbf{B}'$), respectively. Moreover, denote the local CSI, QSI and BSI of EHS node $n$ under system state $\mathbf{S}$ (resp. $\mathbf{S}'$) by $H_{n}$ (resp. $H'_{n}$), $Q_{n}$ (resp. $Q'_{n}$) and $B_{n}$ (resp. $B'_{n}$), respectively.\par

Given $\pi(\Omega)$, the performance measures such as the packet loss rate and average delay for all the EHS nodes can be derived, where $\mathbf{E}^{\pi(\Omega)}[\cdot]$ denotes the expectation operation taken w.r.t. the unique steady-state distribution induced by the given policy $\Omega$.\par

\subsubsection{Packet loss rate}
Packet loss rate is used to quantify the system reliability. The lost packets include (1) erroneously received packets due to deep fading of wireless channel; (2) dropped packets due to data buffer overflow; (3) dropped packets ``dropped" here means failed to be sensed by the sensor) due to lack of sensing energy. The first and second types of lost packets are related to the data communications subsystem, while the third type related to the data sensing subsystem. The first type of lost packets can be measured by BER and guaranteed to be smaller than $\epsilon_{n}$ by the physical layer with an achievable data rate $r(H_{n},p_{n}^{\mathrm{T}})$ given by \eqref{eq1}.\par

In order to measure the second and third types of lost packets, define the packet drop rate $d_{n}$ of EHS node $n$, which can be estimated as
\begin{align}
\label{eq5}
d_{n}=\frac{\mathrm{Average~\#~of~data~units \
dropped \ in \ a \ time \ slot}}{\mathrm{Average~\#~of~data~units \
arrived \ in \ a \ time \ slot}}=1-\frac{\overline{R}_{n}}{\lambda_{n}^{\mathrm{A}}},
\end{align}
\noindent where
\begin{equation}
\label{eq6}
\overline{R}_{n}=\mathbf{E}^{\pi(\Omega)}\left[\min\left[Q_{n},r\left(H_{n},p_{n}^{\mathrm{T}}\right)\right]\right].
\end{equation}
\noindent is the average throughput for any EHS node $n\in\{1,\ldots,N\}$. Note that packets that arrive but not transmitted by a EHS node are either not sensed due to lack of sensing power or dropped due to data buffer overflow.\par

Given a specific BER $\epsilon_{n}$ that can be achieved by EHS node $n$, its packet loss rate can be derived as
\begin{equation}
\label{eq7}
l_{n}=1-(1-\epsilon_{n})^{K}(1-d_{n}),
\end{equation}
\noindent which quantifies the reliability of both the data sensing subsystem and data communications subsystem.\par

\subsubsection{Average delay}
The average delay for any EHS node $n\in\{1,\ldots,N\}$ can be calculated according to Little's Law as
\begin{equation}
\label{eq8}
\overline{D}_{n}=\frac{\overline{Q}_{n}}{\overline{R}_{n}},
\end{equation}
\noindent where
\begin{equation}
\label{eq9}
\overline{Q}_{n}=\mathbf{E}^{\pi(\Omega)}[Q_{n}]
\end{equation}
\noindent is the average queue length of the data buffer of EHS node $n$, and $\overline{R}_{n}$ can be derived from \eqref{eq6}, which equals the effective average arrival rate of packets to EHS node $n$, i.e., the average rate at which the packets successfully enter the data buffer of EHS node $n$.\par

Our objective is to minimize the weighted packet loss rate under the constraint of average delay. Given a BER $\epsilon_{\mathrm{max}}$ that can be achieved by the physical layer, the problem turns into minimizing the weighted packet drop rate according to \eqref{eq7}. Let $D_{\mathrm{max}}$ denote the maximum tolerable average delay for every EHS node, and $\omega_{n} \ (n=1,\ldots,N)$ denote the weight of the packet loss rate of EHS node $n$. Using MDP formalism, the design of optimal scheduling and energy allocation can be formulated as the CMDP problem given in Problem 1.
\newtheorem{problem}{Problem}
\begin{problem}
\begin{equation}
\label{eq11}
\min_{\Omega}\lim_{T\rightarrow\infty}\frac{1}{T}\sum_{t=1}^{T}\mathbf{E}^{\Omega}[g_{0}(\mathbf{S}_{t},\Omega(\mathbf{S}_{t}))]
\end{equation}
\begin{displaymath}
\mathrm{s.t.} \ \lim_{T\rightarrow\infty}\frac{1}{T}\sum_{t=1}^{T}\mathbf{E}^{\Omega}[g_{n}(\mathbf{S}_{t},\Omega(\mathbf{S}_{t}))]\leq 0, \ n=1,\ldots,N,
\end{displaymath}

\noindent where $\mathbf{E}^{\Omega}[\cdot]$ is taken w.r.t the probability measure induced by the policy $\Omega$, and $g_{0}(\mathbf{S}_{t},\Omega(\mathbf{S}_{t}))$ is the reward function related to the optimization objective, $\{g_{n}(\mathbf{S}_{t},\Omega(\mathbf{S}_{t}))\}_{n=1}^{N}$ is a set of reward functions related to the $N$ constraints for average delay. Under any unichain policy\footnote{Since we deal with finite- and countable-state model in this paper, any stationary deterministic policy leads to unichain DTMCs where the transition matrix consists of a single recurrent class \cite{book:Puterman}.} we have $\forall \ n=0,1,\ldots,N$
\begin{displaymath}
\lim_{T\rightarrow\infty}\frac{1}{T}\sum_{t=1}^{T}\mathbf{E}^{\Omega}[g_{n}(\mathbf{S}_{t},\Omega(\mathbf{S}_{t}))]=\mathbf{E}^{\pi(\Omega)}[g_{n}(\mathbf{S},\Omega(\mathbf{S}))].
\end{displaymath}
\noindent Therefore, the reward functions can be derived according to \eqref{eq6}, \eqref{eq8} and \eqref{eq9} as
\begin{equation}
\label{eq12}
g_{0}(\mathbf{S},\mathbf{x})=\sum_{n=1}^{N}\frac{\omega_{n}}{\lambda_{n}}\left[\lambda_{n}-\min\left[Q_{n},r\left(H_{n},p_{n}^{\mathrm{T}}\right)\right]\right],
\end{equation}
\begin{equation}
\label{eq13}
g_{n}(\mathbf{S},\mathbf{x})=Q_{n}-D_{\mathrm{max}}\left[\min\left[Q_{n},r\left(H_{n},p_{n}^{\mathrm{T}}\right)\right]\right], \ n=1,\ldots,N.
\end{equation}
\end{problem}


For any given nonnegative Lagrangian Multipliers (LMs) $\boldsymbol{\eta}$, where $\boldsymbol{\eta}=\{\eta_{n}\}_{n=1}^{N}$, we define the Lagrangian function of Problem 1 as
\begin{equation}
\label{eq14}
L(\Omega,\eta)=\mathbf{E}^{\pi(\Omega)}[g(\mathbf{S},\Omega(\mathbf{S}))]
\end{equation}
\noindent where
\begin{align}
\label{eq15}
g(\mathbf{S},\Omega(\mathbf{S}))=\sum_{n=1}^{N}\omega_{n}+\sum_{n=1}^{N}\eta_{n}Q_{n} -\sum_{n=1}^{N}[\frac{\omega_{n}}{\lambda_{n}}+\eta_{n}D_{\mathrm{max}}]\left[\min\left[Q_{n},r\left(H_{n},p_{n}^{\mathrm{T}}\right)\right]\right]
\end{align}

Therefore, Problem 1 can be decomposed into the following two subproblems:
\begin{displaymath}
\textbf{Subproblem 1-1:  }     G(\boldsymbol{\eta})=\min_{\Omega}L(\Omega,\boldsymbol{\eta}),
\end{displaymath}
\begin{displaymath}
\textbf{Subproblem 1-2:  }  G(\boldsymbol{\eta}^{*})=\max_{\boldsymbol{\eta}}G(\boldsymbol{\eta}).
\end{displaymath}

\section{Energy Allocation and Scheduling Algorithm}
Given the LMs $\boldsymbol{\eta}^{*}$, Subproblem 1-1 is a classical infinite horizon average reward MDP problem, which can be solved by the Bellman's equation with offline value iteration (OVI) algorithm. However, the optimal LMs $\boldsymbol{\eta}^{*}$ in Subproblem 1-2 need to be determined to solve Problem 1. Moreover, the curse of dimensionality problem forbids practical implementation of the brute-force OVI algorithm. In this section, we will first assume that the optimal LMs are given and focus on the solution of Subproblem 1-1 in Section V.A-V.B. Specifically, an equivalent Bellman's equation is constructed in Section V.A as a first step to reduce the state space of the MDP model. Moreover, since the values of $A_{n,t}$ and $E_{n,t}$ are only available at the end of every time slot, while an optimal action has to be determined at the beginning of the time slot, a post-decision state framework is defined in Section V.A to solve this problem. Next, we will use linear value function approximation method to further reduce the state space, and enable the EHS nodes to distributively determine the optimal actions with minimal help from the FC in Section V.B. As a result, Algorithm 1 is proposed at the end of Section V.B which can determine the optimal action assuming that the value functions are given. Then, in Section V.C, we will use an online stochastic learning (OSL) algorithm with two time scales instead of the OVI algorithm to determine the optimal LMs and value functions, so that there is no need to explicitly derive the CSI, packets and EH arrival probability distributions. Finally, Algorithm 2 is proposed at the end of Section V.C as the complete solution to Problem 1.\par

\subsection{Reduced-State Post-Decision Bellman's Equation}
Subproblem 1-1 with given LMs $\boldsymbol{\eta}^{*}$ can be solved by the Bellman's equation, we have $\forall \mathbf{S}\in\mathcal{S}$
\begin{equation}
\label{eq16}
\theta+V(\mathbf{S})=\min_{\Omega(\mathbf{S})} \left\{g(\mathbf{S},\Omega(\mathbf{S}))+\sum_{\mathbf{S}'\in\mathcal{S}}\mathrm{Pr.}[\mathbf{S}'|\mathbf{S},\Omega(\mathbf{S})]V(\mathbf{S}')\right\},
\end{equation}
\noindent where $V(\mathbf{S})$ is the value function representing the average reward obtained following policy $\Omega$ from each state $\mathbf{S}$, while $\theta$ represents the optimal average reward per period for a system in steady-state.\par

As a remark, note that the Bellman's equation \eqref{eq16} represents a series of fixed-point equations, where the number of equations are determined by the number of value functions $V(\mathbf{S})$, which is $|\mathcal{S}|$. Theoretically, the BS can use the brute force value iteration method to offline solve \eqref{eq16} and derive the optimal control policy, in which $|\mathcal{S}|$ value functions need to be stored and the computation complexity is $O(|\mathcal{S}|^{2}|\mathcal{A}|)$ in one iteration, where the number of iterations depends on the convergence speed of the offline value iteration algorithm. Therefore, the offline value iteration algorithm is too complicated to compute due to curse of dimensionality, i.e., the exponential growth of the cardinality of the system state space and the large dimension of the control action space involved. In the rest of this section, we will develop an algorithm with reduced complexity using a series of techniques including equivalent Bellman's equation, post-decision state, linear value approximation, and online stochastic learning.\par

As a first step to reduce the state space of the above MDP, an equivalent Bellman's equation is constructed. We first define the partitioned actions of a policy $\Omega$ as follows.
\newtheorem{definition}{Definition}
\begin{definition}
Given a control policy $\Omega$, we define
\begin{displaymath}
\Omega(\mathbf{Q},\mathbf{B})=\{\Omega(\mathbf{H},\mathbf{Q},\mathbf{B})|\forall \mathbf{H}\}\subseteq \mathcal{A}
\end{displaymath}
\noindent as the collection of $|\mathcal{H}|$ actions, where every action is mapped by policy $\Omega$ from a system state with given QSI $\mathbf{Q}$ and BSI $\mathbf{B}$, and a different realization of CSI $\mathbf{H}\in\mathcal{H}$.
\end{definition}

\newtheorem{lemma}{Lemma}
\begin{lemma}
The control policy obtained by solving the original Bellman's equation \eqref{eq16} is equivalent to the control policy obtained by solving the reduced-state Bellman's equation \eqref{eq17}
\begin{align}
\label{eq17}
&\theta+V\left(\mathbf{Q},\mathbf{B}\right)=\min_{\Omega(\mathbf{Q},\mathbf{B})}\Big\{g(\mathbf{Q},\mathbf{B},\Omega(\mathbf{Q},\mathbf{B})) +\sum_{\mathbf{Q}',\mathbf{B}'}\mathrm{Pr.}[\mathbf{Q}'|\mathbf{Q},\mathbf{B},\Omega(\mathbf{Q},\mathbf{B})] \mathrm{Pr.}[\mathbf{B}'|\mathbf{B},\Omega(\mathbf{Q},\mathbf{B})]\nonumber \\
&V(\mathbf{Q}',\mathbf{B}')\Big\}, \forall \mathbf{Q}\in\mathcal{Q},\mathbf{B}\in\mathcal{B},
\end{align}
\noindent where
\begin{displaymath}
V(\mathbf{Q},\mathbf{B})=\mathbf{E}_{\mathbf{H}}[V(\mathbf{S})|(\mathbf{Q},\mathbf{B})]
=\sum_{\mathbf{H}\in\mathcal{H}}\mathrm{Pr.}[\mathbf{H}]V(\mathbf{H},\mathbf{Q},\mathbf{B})
\end{displaymath}
\noindent is the conditional expectation of value function $V(\mathbf{S})$ taken over the CSI space $\mathcal{H}$ given the QSI $\mathbf{Q}$ and BSI $\mathbf{B}$, while
\begin{displaymath}
g(\mathbf{Q},\mathbf{B},\Omega(\mathbf{Q},\mathbf{B}))=\mathbf{E}_{\mathbf{H}}[g(\mathbf{S},\Omega(\mathbf{S}))|\mathbf{Q},\mathbf{B}],
\end{displaymath}
\begin{displaymath}
\mathrm{Pr.}\left[\mathbf{Q}'|\mathbf{Q},\mathbf{B},\Omega(\mathbf{Q},\mathbf{B})\right]=\mathbf{E}_{\mathbf{H}}\left[
\mathrm{Pr.}\left[\mathbf{Q}'|\mathbf{H},\mathbf{Q},\mathbf{B},\Omega(\mathbf{S})\right]|\mathbf{Q},\mathbf{B}\right],
\end{displaymath}
\begin{displaymath}
\mathrm{Pr.}\left[\mathbf{B}'|\mathbf{B},\Omega(\mathbf{Q},\mathbf{B})\right]=\mathbf{E}_{\mathbf{H}}\left[
\mathrm{Pr.}\left[\mathbf{B}'|\mathbf{B},\Omega(\mathbf{S})\right]|\mathbf{Q},\mathbf{B}\right],
\end{displaymath}
\noindent are conditional expectations of reward function $g(\mathbf{S},\Omega(\mathbf{S}))$ and transition probabilities $\mathrm{Pr.}[\mathbf{Q}'|\mathbf{H},$ $\mathbf{Q},\mathbf{B},\Omega(\mathbf{S})]$, $\mathrm{Pr.}[\mathbf{B}'|\mathbf{B},\Omega(\mathbf{S})]$ taken over the CSI space $\mathcal{H}$ given the QSI $\mathbf{Q}$ and BSI $\mathbf{B}$, respectively.
\end{lemma}

The proof of Lemma 1 is given in Appendix A. As a remark, note that equivalent Bellman's equation \eqref{eq17} also represents a series of fixed-point equations, where the number of equations is determined by the number of value functions $V(\mathbf{Q},\mathbf{B})$, which is $|\mathcal{Q}|\times |\mathcal{B}|$. Therefore, we only need to solve $|\mathcal{Q}|\times |\mathcal{B}|$ instead of $|\mathcal{H}|\times|\mathcal{Q}|\times |\mathcal{B}|$ fixed-point equations with the reduced-state Bellman's equation \eqref{eq17}. In order to solve one such fixed-point equation using value iteration, the R.H.S. of \eqref{eq17} has to be minimized with given value functions $V(\mathbf{Q}',\mathbf{B}')$. For this purpose, the R.H.S. of \eqref{eq17} can be written as
\begin{equation}
\label{eq18}
\min_{\Omega} \sum_{\mathbf{H}\in\mathcal{H}}\mathrm{Pr.}[\mathbf{H}]f(\mathbf{S},\Omega(\mathbf{S})),
\end{equation}
\noindent where
\begin{align}
\label{eq19}
f(\mathbf{S},\Omega(\mathbf{S}))=g(\mathbf{S},\Omega(\mathbf{S})) +\sum_{\mathbf{Q}',\mathbf{B}'}\mathrm{Pr.}[\mathbf{Q}'|\mathbf{H},\mathbf{Q},\mathbf{B},\Omega(\mathbf{S})]\mathrm{Pr.}[\mathbf{B}'|\mathbf{B},\Omega(\mathbf{S})]V(\mathbf{Q}',\mathbf{B}').
\end{align}
\noindent Since \eqref{eq18} is a decoupled objective function w.r.t. different CSI realizations $\mathbf{H}$ with a given QSI $\mathbf{Q}$ and a BSI $\mathbf{B}$, we need to obtain $|\mathcal{H}|$ optimal actions in order to achieve the minimization objective in the R.H.S. of equivalent Bellman equation \eqref{eq17}, where every optimal action is w.r.t. a system state $(\mathbf{H},\mathbf{Q},\mathbf{B})$ with given $\mathbf{Q}$ and $\mathbf{B}$, as well as a different CSI realization $\mathbf{H}\in\mathcal{H}$ that minimizes the value of $f((\mathbf{H},\mathbf{Q},\mathbf{B}),\Omega(\mathbf{H},\mathbf{Q},\mathbf{B}))$. This means that the control policy obtained by solving \eqref{eq17} is based on the system state $\mathbf{S}$ in addition to the queue state $\mathbf{Q}$ and battery energy state $\mathbf{B}$, although the value function $V(\mathbf{Q},\mathbf{B})$ does not depend on the CSI $\mathbf{H}$. Also note that $V(\mathbf{Q},\mathbf{B})$ is affected by the CSI distribution.\par


In order to derive an optimal action under every system state to minimize \eqref{eq19}, the knowledge of the transition probabilities $\mathrm{Pr.}[\mathbf{Q}'|\mathbf{H},\mathbf{Q},\mathbf{B},\Omega(\mathbf{S})]$ and $\mathrm{Pr.}[\mathbf{B}'|\mathbf{B},\Omega(\mathbf{S})]$ are required, which in turn require the distributions of the EH arrival process $\mathrm{Pr.}(\mathbf{E})=\prod_{n=1}^{N}f_{E}(E_{n})$ and packets arrival process $\mathrm{Pr.}(\mathbf{A})=\prod_{n=1}^{N}f_{A}(A_{n})$ that are unknown to EHS nodes. Moreover, the values of $A_{n,t}$ and $E_{n,t}$ are only available at the end of time slot $t$, while an optimal action has to be determined at the beginning of time slot $t$. In order to address this limitation, we define the post-decision reduced state \cite{EJOR:Magirou} to be the virtual QSI and BSI immediately after making an action but before the new harvested energy and new sensed data arrive. Let $(\mathbf{Q},\mathbf{B})$ be the reduced state at the beginning of time slot $t$ (also called the pre-decision reduced state), and the EHS nodes make an action that consumes energy $\mathbf{p}^{\mathrm{T}}$ during time slot $t$, then the post-decision reduced state is denoted by $(\widetilde{\mathbf{Q}}',\widetilde{\mathbf{B}}')=\left(\max[0,\mathbf{Q}-\mathbf{r}(\mathbf{H},\mathbf{p}^{\mathrm{T}})],\mathbf{B}-\mathbf{p}^{\mathrm{T}}\right)$. Let $\mathbf{A}'=\{A_{n}'|n=1,\ldots,N\}$ and $\mathbf{E}'=\{E_{n}'|n=1,\ldots,N\}$ be two vectors representing the packets and EH arrivals for all the EHS nodes during time slot $t$, respectively. Then the system reaches the next actual reduced state, i.e., the pre-decision reduced state at time slot $t+1$ $(\mathbf{Q}',\mathbf{B}')=(\min[Q_{\mathrm{\mathrm{max}}}\mathbf{e},\max[0,\mathbf{Q}-\mathbf{r}(\mathbf{H},\mathbf{p}^{\mathrm{T}})]+\min[\mathbf{A}',\gamma(\mathbf{B}-\mathbf{p}^{\mathrm{T}})],\min[B_{\mathrm{\mathrm{max}}}\mathbf{e},\max[0,\mathbf{B}-\mathbf{p}^{\mathrm{T}}-\mathbf{A}'/\gamma]+\mathbf{E}'])$. On the other hand, it can be deduced that the pre-decision reduced state $(\mathbf{Q},\mathbf{B})$ at time slot $t$ is reached from the post-decision reduced state $(\widetilde{\mathbf{Q}},\widetilde{\mathbf{B}})$ at time slot $t-1$ as $(\mathbf{Q},\mathbf{B})=(\min[Q_{\mathrm{max}}\mathbf{e},\widetilde{\mathbf{Q}}+\min[\mathbf{A},\gamma\widetilde{\mathbf{B}}]],\min[B_{\mathrm{max}}\mathbf{e},\max[0,\widetilde{\mathbf{B}}-\mathbf{A}/\gamma]+\mathbf{E}])$, where $\mathbf{A}=\{A_{n}|n=1,\ldots,N\}$ and $\mathbf{E}=\{E_{n}|n=1,\ldots,N\}$ are the vectors representing the packets and EH arrivals for all the EHS nodes during time slot $t-1$, respectively.\par

\newtheorem{lemma2}[lemma]{Lemma}
\begin{lemma2}
The control policy obtained by solving the reduced-state Bellman's equation \eqref{eq17} is equivalent to the post-decision Bellman's equation \eqref{eq20} :

\begin{align}
\label{eq20}
&\theta+V(\widetilde{\mathbf{Q}},\widetilde{\mathbf{B}})=\sum_{\mathbf{Q},\mathbf{B}}\mathrm{Pr.}[\mathbf{Q}|\widetilde{\mathbf{Q}},\widetilde{\mathbf{B}}]\mathrm{Pr.}[\mathbf{B}|\widetilde{\mathbf{B}}]\min_{\Omega(\mathbf{Q},\mathbf{B})}\Big\{g(\mathbf{Q},\mathbf{B}, \Omega(\mathbf{Q},\mathbf{B}))+\sum_{\widetilde{\mathbf{Q}}',\widetilde{\mathbf{B}}'}\mathrm{Pr.}[\widetilde{\mathbf{Q}}'|\mathbf{Q},\Omega(\mathbf{Q},\mathbf{B})]\nonumber \\
&\mathrm{Pr.}[\widetilde{\mathbf{B}}'|\mathbf{B},\Omega(\mathbf{Q},\mathbf{B})]V(\widetilde{\mathbf{Q}}',\widetilde{\mathbf{B}}')\Big\}, \ \forall \widetilde{\mathbf{Q}}\in\mathcal{Q},\widetilde{\mathbf{B}}\in\mathcal{B},
\end{align}

\noindent where
\begin{displaymath}
V(\widetilde{\mathbf{Q}},\widetilde{\mathbf{B}})=\mathbf{E}_{\mathbf{Q}|\widetilde{\mathbf{Q}},\mathbf{B}|\widetilde{\mathbf{B}}}[V(\mathbf{Q},\mathbf{B})]
\end{displaymath}
\noindent is the expectation taken over all the pre-decision reduced states that can be reached from the post-decision reduced state $(\widetilde{\mathbf{Q}},\widetilde{\mathbf{B}})$.
\end{lemma2}

The proof of Lemma 2 is given in Appendix B. As a remark, note that post-decision Bellman's equation \eqref{eq20} also represents $|\mathcal{Q}|\times |\mathcal{B}|$ fixed-point equations. Since the transition probabilities $\mathrm{Pr.}[\mathbf{Q}|\widetilde{\mathbf{Q}},\widetilde{\mathbf{B}}]$ and $\mathrm{Pr.}[\mathbf{B}|\widetilde{\mathbf{B}}]$ from given post-decision reduced state $(\widetilde{\mathbf{Q}},\widetilde{\mathbf{B}})$ to different pre-decision reduced state $(\mathbf{Q},\mathbf{B})$ only depend on $\mathbf{A}$ and $\mathbf{E}$ as discussed above, the R.H.S. of \eqref{eq20} is a decoupled objective function w.r.t. different packets and EH arrivals with a given post-decision reduced state $(\widetilde{\mathbf{Q}},\widetilde{\mathbf{B}})$. Since the optimal action is determined at the beginning of time slot $t$, when the values of $\mathbf{A}$ and $\mathbf{E}$ (the number of arrived packets and harvest energy units during time slot $t-1$) are already known, we only need to minimize the following function $\widetilde{f}(\mathbf{S})$ in order to minimize $f(\mathbf{S})$ as defined in \eqref{eq19} to derive the optimal action under system state $\mathbf{S}$, where

\begin{align}
\label{eq21}
\widetilde{f}(\mathbf{S})=g(\mathbf{S},\Omega(\mathbf{S}))+ \sum_{\widetilde{\mathbf{Q}}',\widetilde{\mathbf{B}}'} \mathrm{Pr.}[\widetilde{\mathbf{Q}}'|\mathbf{H},\mathbf{Q},\Omega(\mathbf{S})]\mathrm{Pr.}[\widetilde{\mathbf{B}}'|\mathbf{B},\Omega(\mathbf{S})]V(\widetilde{\mathbf{Q}}',\widetilde{\mathbf{B}}').
\end{align}

\noindent Note that the distributions $\mathrm{Pr.}(\mathbf{A})$ and $\mathrm{Pr.}(\mathbf{E})$ are no longer needed in order to minimize $\widetilde{f}(\mathbf{S})$. When obtaining the optimal action using \eqref{eq21}, we assume that the value functions $V(\widetilde{\mathbf{Q}},\widetilde{\mathbf{B}}),\widetilde{\mathbf{Q}}\in\mathcal{Q},\widetilde{\mathbf{B}}\in\mathcal{B}$ are given. Although $V(\widetilde{\mathbf{Q}},\widetilde{\mathbf{B}})$ depends on $\mathrm{Pr.}(\mathbf{A})$ and $\mathrm{Pr.}(\mathbf{E})$, we will show in Section V.C that its value can be obtained using online stochastic learning algorithm without explicitly deriving the arrival probability distributions. Therefore, $\mathrm{Pr.}(\mathbf{A})$ and $\mathrm{Pr.}(\mathbf{E})$ are not needed for solving the MDP problem.\par

\subsection{Linear Value Function Approximation}
First, we define the per-node reward function as
\begin{align}
\label{eq22}
g_{n}(\mathbf{S}_{n},\Omega(\mathbf{S}))=\omega_{n}+\eta_{n}Q_{n} -[\frac{\omega_{n}}{\lambda_{n}}+\eta_{n}D_{\mathrm{max}}]\left[\min\left[Q_{n},r\left(H_{n},p_{n}^{\mathrm{T}}\right)\right]\right].
\end{align}
\noindent Thus, the overall reward function is given by $g(\mathbf{S},\Omega(\mathbf{S}))=\sum_{n=1}^{N}g_{n}(\mathbf{S}_{n},\Omega(\mathbf{S}))$ according to \eqref{eq15}.

Next, the linear approximation architecture for the value function $V(\widetilde{\mathbf{Q}},\widetilde{\mathbf{B}})$ is given by
\begin{align}
\label{eq23}
V(\widetilde{\mathbf{Q}},\widetilde{\mathbf{B}})=V(\{\widetilde{Q}_{n},\widetilde{B}_{n}\}_{n=1}^{N})\approx\sum_{n=1}^{N}\widetilde{V}_{n}(\widetilde{Q}_{n},\widetilde{B}_{n})=\sum_{n=1}^{N}\sum_{q=0}^{Q_{\mathrm{max}}}\sum_{b=0}^{B_{\mathrm{max}}}\mathbf{I}[\widetilde{Q}_{n}=q,\widetilde{B}_{n}=b]\widetilde{V}_{n}(q,b).
\end{align}
\noindent We refer to $\widetilde{V}_{n}(\widetilde{Q}_{n},\widetilde{B}_{n}) \textrm{ or } \widetilde{V}_{n}(q,b)$ as per-node value function and $V(\widetilde{\mathbf{Q}},\widetilde{\mathbf{B}})$ as global value function in the rest of the paper. Therefore, $\widetilde{\mathbf{V}}_{n}=\left[\widetilde{V}_{n}(0,0),\ldots,\widetilde{V}_{n}(Q_{\mathrm{max}},B_{\mathrm{max}})\right]$ is the per-node value function vector for EHS node $n$, $\widetilde{\mathbf{V}}=\left[\widetilde{\mathbf{V}}_{n}|n=1,\ldots,N\right]^{T}$ is the per-node value function vector for all the EHS nodes in the network. Similarly, define the global value function vector as $\mathbf{V}=\left[V(\widetilde{\mathbf{Q}},\widetilde{\mathbf{B}})|\widetilde{\mathbf{Q}}\in\mathcal{Q},\widetilde{\mathbf{B}}\in\mathcal{B}\right]^{T}$.

As a remark, note that the number of global value functions is $|\mathcal{Q}|\times |\mathcal{B}|=[(Q_{\mathrm{max}}+1)(B_{\mathrm{max}}+1)]^{N}$ in total, which grows exponentially with the number of nodes. On the other hand, the number of per-node value functions is $(Q_{\mathrm{max}}+1)(B_{\mathrm{max}}+1)\times N$ in total, which grows linearly with the number of nodes. Therefore, we can represent the $[(Q_{\mathrm{max}}+1)(B_{\mathrm{max}}+1)]^{N}$ global value functions with $(Q_{\mathrm{max}}+1)(B_{\mathrm{max}}+1)\times N$ per-node value functions by the linear approximation architecture.\par


We assume that every EHS node $n$ maintains its local per node value function $\widetilde{\mathbf{V}}_{n}$ and LM $\eta_{n}$. From \eqref{eq21}, the key step in deriving the optimal control actions is to obtain the global value function vector $\mathbf{V}$. With linear value function approximation, we only need to obtain the per-node value function vector $\widetilde{\mathbf{V}}$. To illustrate the structure of our algorithm, we first assume we could obtain the per-node value functions via some means (e.g., via offline value iteration) and focus on deriving the optimal action under every system state to minimize the value of $\widetilde{f}(\mathbf{S})$. The optimal control action is given by the following Subproblem 1-1(a).\par

\newtheorem*{proaux}{Subproblem \protect\pronumber}
\newenvironment{pro}[1]{\def\pronumber{#1}\proaux}{\endproaux}
\begin{pro}{1-1(a)}
For given per-node value functions $\widetilde{\mathbf{V}}$ and LMs $\boldsymbol{\eta}$, find the optimal action $\Omega^{*}(\mathbf{H},\mathbf{Q},\mathbf{B})$ for system state $(\mathbf{H},\mathbf{Q},\mathbf{B})$ that minimizes the value of $\widetilde{f}(\mathbf{H},\mathbf{Q},\mathbf{B})$, which can be written as follows
\begin{align}
\Omega^{*}(\mathbf{S})
=&\arg\min_{\Omega}\sum_{n=1}^{N}g_{n}\left(\mathbf{S}_{n},\Omega(\mathbf{S})\right)  +\sum_{\widetilde{\mathbf{Q}}',\widetilde{\mathbf{B}}'}\left\{\prod_{n=1}^{N}\mathrm{Pr.}\left[(\widetilde{Q}_{n}',\widetilde{B}_{n}')|\mathbf{S}_{n},\Omega(\mathbf{S})\right]\sum_{n=1}^{N}\widetilde{V}_{n}(\widetilde{Q}_{n}',\widetilde{B}_{n}')\right\} \label{eq24} \\
=&\arg\min_{\Omega}\sum_{n=1}^{N}\left\{g_{n}\left(\mathbf{S}_{n},p_{n}^{T}\right)+\widetilde{V}_{n}\left(Q_{n}\left(r(H_{n},p_{n}^{T})\right),B_{n}\left(p_{n}^{T}\right)\right)\right\}, \label{eq25}
\end{align}

\noindent where the QSI $Q_{n}\left(r(H_{n},p_{n}^{T})\right)$ is defined by the following equation
\begin{equation}
\label{eq26}
Q_{n}\left(r(H_{n},p_{n}^{T})\right)=\max[0,Q_{n}-r(H_{n},p_{n}^{\mathrm{T}})],
\end{equation}
\noindent and the BSI $B_{n}(p_{n}^{T})$ is defined by the following equation
\begin{equation}
\label{eq27}
B_{n}(p_{n}^{\mathrm{T}})=B_{n}-p_{n}^{\mathrm{T}}.
\end{equation}
\end{pro}
\noindent Step \eqref{eq24} follows from the linear value approximation structure in \eqref{eq23}. Step \eqref{eq25} holds because of the queues dynamics and battery energy dynamics from pre-decision reduced state to post-decision reduced state. \par

\newtheorem{algorithm}{Algorithm}
\begin{algorithm}[solution to Subproblem 1-1(a)]
Given per-node value functions $\widetilde{\mathbf{V}}$ and LMs $\boldsymbol{\eta}$, the optimal action in subproblem 1-1(a) for every system state $\mathbf{S}$ is determined as
\begin{equation}
\label{eq31}
x_{n}^{*}=\left\{
\begin{array}{ll}
1, & \textrm{if }n=\arg\max_{n'=1}^{N}bid_{n'}, \\
0, & \textrm{otherwise.}
\end{array} \right.
\end{equation}
\noindent where
\begin{align}
\label{eq30}
bid_{n}=\Big\{[\frac{\omega_{n}}{\lambda_{n}}+\eta_{n}D_{\max}+\big(\widetilde{V}_{n}^{(Q)}(Q_{n},B_{n})\big)']r(H_{n},p_{n}^{\mathrm{T}}) +p_{n}^{\mathrm{T}}\big(\widetilde{V}_{n}^{(B)}(Q_{n},B_{n})\big)' \Big\}.
\end{align}
\begin{align}
\label{eq32}
p_{n}^{\mathrm{T}*}=x_{n}\min[B_{n},\frac{N_{0}\tau W (2^{\frac{Q_{n}K}{\tau W}}-1)}{\xi_{n}H_{n}},\max[0,\frac{(\frac{\omega_{n}}{\lambda_{n}}+\eta_{n}D_{\max}+\big(\widetilde{V}_{n}^{(Q)}(Q_{n},B_{n})\big)')\tau W}{-K\big(\widetilde{V}_{n}^{(B)}(Q_{n},B_{n})\big)'\ln2}-\frac{N_{0}W\tau}{\xi_{n}H_{n}}]].
\end{align}
\end{algorithm}
%

\newtheorem{remark}{Remark}
\begin{remark}[Implementation and Complexity of Algorithm 1]
Due to linear value function approximation, only $(Q_{\mathrm{max}}+1)(B_{\mathrm{max}}+1)\times N$ per-node value functions need to be stored instead of $[(Q_{\mathrm{max}}+1)(B_{\mathrm{max}}+1)]^{N}$ global value functions. Specifically, each EHS node only need to store $(Q_{\mathrm{max}}+1)(B_{\mathrm{max}}+1)$ local per node value functions. At the beginning of every time slot $t$, each EHS node $n$ observes its local system state and determines its optimal transmission power $p_{n,t}^{\mathrm{T}*}$ if scheduled according to \eqref{eq32}. Then, it submits its bid, $bid_{n}$, to FC, which is calculated by taking $p_{n,t}^{\mathrm{T}*}$ into \eqref{eq30}. The FC makes a decision about scheduling the EHS node with the largest $bid_{n}$ according to \eqref{eq31}, and broadcasts the scheduling action $x_{n,t}$ to all EHS nodes. Then the scheduled EHS node $n$ transmits with power $p_{n,t}^{\mathrm{T}*}$. Therefore, the overall computational complexity of Algorithm 1 is $\mathcal{O}(2N)$.
\end{remark}

The proof of Algorithm 1 is given in Appendix C. In the above discussion, we assume that the per-node value function vector $\widetilde{\mathbf{V}}$ is already known in Subproblem 1-1(a) and propose Algorithm 1 in order to derive the optimal control action under every system state. However, we still need to determine $\widetilde{\mathbf{V}}$ in order to solve Subproblem 1-1. Let $\widetilde{V}_{n}(0,0)=0$, $\forall n=1,\ldots,N$. According to the linear approximation architecture, among the $[(Q_{\mathrm{max}}+1)(B_{\mathrm{max}}+1)]^{N}$ global value functions, there are $[(Q_{\mathrm{max}}+1)(B_{\mathrm{max}}+1)-1]\times N$ global value functions that equal to the $[(Q_{\mathrm{max}}+1)(B_{\mathrm{max}}+1)-1]\times N$ per-node value functions $\{\widetilde{V}_{n}(q,b)|\forall q\in\{0,\ldots,Q_{\mathrm{max}}\}, b\in\{0,\ldots,B_{\mathrm{max}}\},q\times b\neq 0\}_{n=1}^{N}$. We refer the system states of these global value functions as \emph{representative states}, and they share the same characteristics that only one node has non-empty data buffer and/or battery energy buffer while both buffers of all the other nodes are empty. The set of representative states $(\widetilde{\mathcal{Q}},\widetilde{\mathcal{B}})_{\mathrm{R}}$ is defined as
\begin{displaymath}
(\widetilde{\mathcal{Q}},\widetilde{\mathcal{B}})_{\mathrm{R}}=\Big\{\left(\widetilde{\mathbf{Q}}_{n}^{(q)},\widetilde{\mathbf{B}}_{n}^{(b)}\right)\mid
\forall q\in\{0,\ldots,Q_{\mathrm{max}}\}, b\in\{0,\ldots,B_{\mathrm{max}}\},q\times b\neq 0\}\Big\}_{n=1}^{N},
\end{displaymath}
\noindent where $\widetilde{\mathbf{Q}}_{n}^{(q)}=\{\widetilde{Q}_{n}=q,\widetilde{Q}_{n'}=0|n'=1,\ldots,N,n'\neq n\}$ denotes the global QSI with $\widetilde{Q}_{n}=q\in\{1,\ldots,Q_{\mathrm{max}}\}$ for EHS node $n$ and $\widetilde{Q}_{n'}=0$ for all the other EHS nodes $n'\neq n$, and $\widetilde{\mathbf{B}}_{n}^{(b)}=\{\widetilde{B}_{n}=b,\widetilde{B}_{n'}=0|n'=1,\ldots,N,n'\neq n\}$ denotes the global post-decision BSI with $\widetilde{B}_{n}=b\in\{1,\ldots,B_{\mathrm{max}}\}$ for EHS node $n$  and $\widetilde{B}_{n'}=0$ for all the other EHS nodes $n'\neq n$. It is worth noting that the state $(\widetilde{\mathbf{Q}}=\mathbf{0},\widetilde{\mathbf{B}}=\mathbf{0})$ is not a representative state. We refer to it as the reference state. Therefore, given the solution of Subproblem 1-1(a), we still need to solve the following Subproblem 1-1(b) in order to solve Subproblem 1-1.
\begin{pro}{1-1(b)}
Derive the per-node value functions $\widetilde{\mathbf{V}}$ that satisfy the following equivalent post-decision Bellman's equation under every representative state $(\widetilde{\mathbf{Q}}_{n}^{(q)},\widetilde{\mathbf{B}}_{n}^{(b)})=(\widetilde{\mathbf{Q}},\widetilde{\mathbf{B}}),\forall \left(\widetilde{\mathbf{Q}}_{n}^{(q)},\widetilde{\mathbf{B}}_{n}^{(b)}\right)\in(\widetilde{\mathcal{Q}},\widetilde{\mathcal{B}})_{\mathrm{R}}$
\begin{align}
\label{eq33}
&\theta+\widetilde{V}_{n}(q,b)=\sum_{\mathbf{Q},\mathbf{B}}\mathrm{Pr.}[\mathbf{Q}|\widetilde{\mathbf{Q}},\widetilde{\mathbf{B}}]\mathrm{Pr.}[\mathbf{B}|\widetilde{\mathbf{B}}] \min_{\Omega}\Big\{g_{n}(Q_{n},B_{n},\Omega(\mathbf{Q},\mathbf{B}))\nonumber \\
&+\sum_{n=1}^{N}\sum_{\widetilde{Q}'_{n}}\mathrm{Pr.}[\widetilde{Q}'_{n}|Q_{n},B_{n},\Omega(\mathbf{Q},\mathbf{B})]\sum_{\widetilde{B}'_{n}}\mathrm{Pr.}[\widetilde{B}'_{n}|B_{n},\Omega(\mathbf{Q},\mathbf{B})]V(\widetilde{Q}'_{n},\widetilde{B}'_{n})\Big\}.
\end{align}

\end{pro}

\subsection{Online Stochastic Learning}
Instead of solving the reduced-state post-decision Bellman's equation on the representative states \eqref{eq33} using offline value iteration, we will estimate $\widetilde{\mathbf{V}}$ using online stochastic learning algorithm. In this way, we can solve the Bellman's equation without the need of explicitly deriving the CSI, packets and EH arrival probability distributions $\mathrm{Pr.}[\mathbf{H}]$, $\mathrm{Pr.}[\mathbf{A}]$ and $\mathrm{Pr.}[\mathbf{E}]$ in order to calculate the ``conditional reward" $g_{n}(Q_{n},B_{n},\Omega(\mathbf{Q},\mathbf{B}))$ and ``conditional transition probability" $\mathrm{Pr.}[\widetilde{Q}'_{n}|Q_{n},B_{n},\Omega(\mathbf{Q},\mathbf{B})]$ and $\mathrm{Pr.}[\widetilde{B}'_{n}|B_{n},\Omega(\mathbf{Q},\mathbf{B})]$ and transition probabilities $\mathrm{Pr.}[\mathbf{Q}|\widetilde{\mathbf{Q}},\widetilde{\mathbf{B}}]$ and $\mathrm{Pr.}[\mathbf{B}|\widetilde{\mathbf{B}}]$ in \eqref{eq33}.\par

The distributed online iterative algorithm (Algorithm 2) is given as follows, which simultaneously solves Subproblem 1-1(b) in deriving per-node value functions and Subproblem 1-2 in deriving LMs $\boldsymbol{\eta}$. \par

\newtheorem{algorithm2}[algorithm]{Algorithm}
\begin{algorithm2}[solution to Subproblem 1-1(b) and Subproblem 1-2]
At the end of time slot $t$, based on the observed system state $\mathbf{S}_{t}$, the post-decision reduced state $(\widetilde{\mathbf{Q}}_{t},\widetilde{\mathbf{B}}_{t})$ at time slot $t-1$ and the optimal action $(\mathbf{x}_{t},\mathbf{p}_{t}^{T})$, the per-node value functions $\widetilde{\mathbf{V}}_{t}$ and LMs $\boldsymbol{\eta}_{t}$ can be updated to $\widetilde{\mathbf{V}}_{t+1}$ and $\boldsymbol{\eta}_{t+1}$ using the update function
\begin{align}
\label{eq34}
\widetilde{V}_{n,t+1}(q,b)=  \left  \{ \begin{array}{lll} \left(1-\varepsilon_{\tau_{n}(q,b,t)}^{v}\right)\widetilde{V}_{n,t}(q,b)+\varepsilon_{\tau_{n}(q,b,t)}^{v}\Delta\widetilde{V}_{n,t}(q,b),
\textrm{if} \ (\widetilde{\mathbf{Q}}_{t},\widetilde{\mathbf{B}}_{t})=(\widetilde{\mathbf{Q}}_{n}^{(q)},\widetilde{\mathbf{B}}_{n}^{(b)}),\\
\widetilde{V}_{n,t}(q,b), \ \ \textrm{if} \ (\widetilde{\mathbf{Q}}_{t},\widetilde{\mathbf{B}}_{t})\neq(\widetilde{\mathbf{Q}}_{n}^{(q)},\widetilde{\mathbf{B}}_{n}^{(b)}),
\end{array} \right.
\end{align}

\noindent where $\varepsilon_{\tau_{n}(q,b,t)}^{v}=\sum_{t'=0}^{t}\mathbf{I}\left[(\widetilde{\mathbf{Q}}_{t'},\widetilde{\mathbf{B}}_{t'})=(\widetilde{\mathbf{Q}}_{n}^{(q)},\widetilde{\mathbf{B}}_{n}^{(b)})\right]$ and
\begin{align*}
&\Delta\widetilde{V}_{n,t}(q,b)=\eta_{n,t}(q+\min[A_{n,t-1},\gamma b])-[\frac{\omega_{n}}{\lambda_{n}} +\eta_{n}D_{\max}]r\left(H_{n,t},p_{n,t}^{T}\right) +\widetilde{V}_{n}\big(\min[Q_{\mathrm{max}},q+\min \\&\gamma [A_{n,t-1},b]]-r(H_{n,t},p_{n,t}^{T}),\min[B_{\mathrm{max}},\max[0,b-A_{n,t-1}/\gamma]+E_{n,t-1}]-p_{n,t}^{T}\big) -\widetilde{V}_{n}\left(0,E_{n,\bar{t}-1}\right),
\end{align*}
\noindent where $\bar{t}=\sup\{t|(\widetilde{\mathbf{Q}}_{t},\widetilde{\mathbf{B}}_{t})=(\mathbf{0},\mathbf{0})\}$.\par

Moreover, the LMs $\eta_{n,t}$ of every node $n$ can be updated at the end of time slot $t$ to $\eta_{n,t+1}$ using the following function
\begin{equation}
\label{eq35}
\eta_{n,t+1}=\eta_{n,t}+\varepsilon_{t}^{\eta }Q_{n}-D_{\mathrm{max}}r\left(H_{n,t},p_{n,t}^{T}\right).
\end{equation}

In the above equations, $\left(\{\varepsilon_{t}^{v}\},\{\varepsilon_{t}^{\eta}\}\right)$ are the sequences of step sizes, which satisfy
\begin{displaymath}
\sum_{t=0}^{\infty}\varepsilon_{t}^{v}=\sum_{t=0}^{\infty}\varepsilon_{t}^{\eta}=\infty, \ \varepsilon_{t}^{v},\varepsilon_{t}^{\eta}>0, \ \lim_{t\rightarrow\infty}\varepsilon_{t}^{v}=\lim_{t\rightarrow\infty}\varepsilon_{t}^{\eta}=0,
\end{displaymath}
\begin{displaymath}
\sum_{t=0}^{\infty}\left[\left(\varepsilon_{t}^{v}\right)^{2}+\left(\varepsilon_{t}^{\eta}\right)^{2}\right]<\infty, \textrm{ and } \lim_{t\rightarrow\infty}\frac{\varepsilon_{t}^{\eta}}{\varepsilon_{t}^{v}}=0.
\end{displaymath}
\end{algorithm2}

\newtheorem{remark2}[remark]{Remark}
\begin{remark2}[Implementation and Complexity of Algorithm 2]
In Algorithm 2, every EHS node can locally update its own per-node value function and LM. Note that $\widetilde{V}_{n,t}(q,b)$ is only updated to a different value at any time slot $t$ when the global post-decision reduced state $(\widetilde{\mathbf{Q}}_{t},\widetilde{\mathbf{B}}_{t})$ is the representative state $(\widetilde{\mathbf{Q}}_{n}^{(q)},\widetilde{\mathbf{B}}_{n}^{(b)})$ according to \eqref{eq34}. This implies that at most one per-node value function shall be updated to a different value at any time slot with computational complexity $\mathcal{O}(N)$, while all the other per-node value functions remain the same. In order to determine whether the current global state is one of the representative states, the FC maintains a bit map with each bit indicating whether both the data buffer and battery energy buffer of an EHS node are empty or not. Every EHS node needs to send an empty flag whenever its buffer status changes from empty to non-empty or vice versa. This is event-triggered and does not need to happen every time slot. If the global state at time slot $t$ $(\widetilde{\mathbf{Q}}_{t},\widetilde{\mathbf{B}}_{t})$ is in representative state $(\widetilde{\mathbf{Q}}_{n}^{(q)},\widetilde{\mathbf{B}}_{n}^{(b)})$, the FC notifies EHS node $n$ using a RS flag.
\end{remark2}

\begin{figure}
\centering
\includegraphics[width=0.6\textwidth]{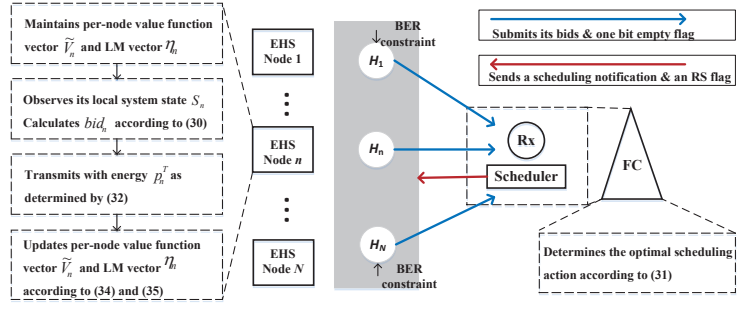}
\caption{The implementation flow of the proposed AMDP+OSL algorithm.}
\label{OSL}
\end{figure}

Algorithm 1 and Algorithm 2 together form the complete solution to Problem 1. We refer to the overall algorithm with approximate MDP
and online stochastic learning as AMDP+OSL algorithm. Fig.\ref{OSL} shows the implementation flow of the proposed AMDP+OSL algorithm with detail steps illustrated as follows:
\begin{itemize}
  \item Step 1 (\textbf{Initialization}): Every EHS node initiates the per-node value function vector $\mathbf{\widetilde{V}}_{n}^{0}$ and LM vector $\eta_{n}^{0}$ that it maintains. The superscript denotes the index of time slot.
  \item Step 2 (\textbf{Calculate Control Action}): At the beginning of the $t$th time slot ($t=1,2,\ldots$), every EHS node observes its local system state $\mathbf{S}_{n,t}$, calculates $bid_{n}$ according to \eqref{eq30} and submits its bid and potentially one bit empty flag to the FC. The FC determines the optimal scheduling action according to \eqref{eq31}. Then the FC sends a scheduling notification to the scheduled EHS node and also an RS flag if the current global state is in the corresponding representative state. The scheduled EHS node transmits with energy as determined by \eqref{eq32}.
  \item Step 3 (\textbf{Update of Per-Node Value Function}): At the end of the $t$th time slot ($t=1,2,\ldots$), with the observed local system state and the RS flag, every EHS node updates the per-node value function vector $\mathbf{\widetilde{V}}_{n,t}$ and LM $\eta_{n,t}$ it maintains according to \eqref{eq34} and \eqref{eq35}. Set $t:=t+1$ and go to Step 2.
\end{itemize}

Note that the computational complexity of AMDP+OSL algorithm at each time slot is the sum of those of Algorithm 1 and Algorithm 2, which is at most $\mathcal{O}(3N)$ and grows linearly with the number of nodes, instead of $[(Q_{\mathrm{max}}+1)(B_{\mathrm{max}}+1)]^{2N}|\mathcal{H}|^{2}|\mathcal{A}|$ with offline value iteration method.\par

\subsection{Convergence Analysis}
In this section, we shall establish technical conditions for the almost-sure convergence of the online stochastic learning algorithm (Algorithm 2). Recall that the purpose of Algorithm 2 is to iteratively derive the per-node value function vector $\widetilde{\mathbf{V}}$ in subproblem 1-1(b) and LMs $\boldsymbol{\eta}$ in subproblem 1-2, so that Problem 1 can be solved. Given $\boldsymbol{\eta}*$, subproblem 1-1 is an unconstraint MDP problem, so learning algorithms in \cite{SIAM:Borkar} apply to update $\widetilde{\mathbf{V}}$. This is precisely what we do in Algorithm 2. But then one has to learn the correct $\boldsymbol{\eta}^{*}$. We do this by a gradient ascent in the dual (i.e., Lagrange multiplier) space in view of subproblem 1-2. Since we have two different step size sequences $\{\epsilon_{t}^{v}\}$ , $\{\varepsilon_{t}^{\eta}\}$, and $\{\varepsilon_{t}^{\eta}\}=\mathbf{o}(\{\varepsilon_{t}^{v}\})$, the LM's update is carried out simultaneously with the per-node value function's update but over a slower timescale. Here we are using the formalism of two timescale stochastic approximation from \cite{SystControl:Borkar}. During the per-node value functions' update (timescale I), we have $\mathbf{\eta}_{n,t+1}-\mathbf{\eta}_{n,t}=\mathcal{O}(\{\varepsilon_{t}^{\eta}\})=\mathbf{o}(\{\varepsilon_{t}^{v}\})$, $\forall c\in\mathcal{C}$ and hence the LMs appear to be quasi-static during the per-node value functions' update in \eqref{eq34}. On the other hand, since the per-node value functions will be updated much faster than the LMs due to $\frac{\varepsilon_{t}^{\eta}}{\varepsilon_{t}^{v}}\rightarrow 0$, during the LMs' update in \eqref{eq35} (timescale II), the `primal' minimization carried out by the learning algorithm for MDPs in \eqref{eq34} is seen as having essentially equilibrated. Therefore, we will give the convergence of per-node value function over timescale I and LMs over timescale II in Lemma 3 and Lemma 4, respectively.\par

Before Lemma 3 is given, we first write the relationship between the global value function vector $\mathbf{V}$ and per-node value function vector $\widetilde{\mathbf{V}}$ in matrix form for easy of notation:
\begin{displaymath}
\mathbf{V}=\mathbf{M}\widetilde{\mathbf{V}} \textrm{      and      } \widetilde{\mathbf{V}}=\mathbf{M}^{\dag}\mathbf{V},
\end{displaymath}
\noindent where $\mathbf{M}\in\mathbb{R}^{(Q_{\mathrm{\mathrm{max}}}+1)(B_{\mathrm{\mathrm{max}}}+1)^{N}\times(Q_{\mathrm{\mathrm{max}}}+1)(B_{\mathrm{\mathrm{max}}}+1)N}$ with the $z$th row $(z=1,\ldots,(Q_{\mathrm{max}}+1)(B_{\mathrm{max}}+1)^{N})$ equals to $\mathbf{F}^{T}(\widetilde{\mathbf{Q}}^{(z)},\widetilde{\mathbf{B}}^{(z)})$, and $(\widetilde{\mathbf{Q}}^{(z)},\widetilde{\mathbf{B}}^{(z)})$ is the $z$-th reduced state in the state space $\mathcal{Q}\times\mathcal{B}$. Therefore, the first equation above follows directly from \eqref{eq24}. The second equation, on the other hand, uses the matrix $\mathbf{M}^{\dag}\in\mathbb{R}^{(Q_{\mathrm{max}}+1)(B_{\mathrm{max}}+1)N\times(Q_{\mathrm{max}}+1)(B_{\mathrm{max}}+1)^{N}}$ to select $(Q_{\mathrm{max}}+1)(B_{\mathrm{max}}+1)N$ elements from $\mathbf{V}$ which correspond to the representative states. Specifically, $\mathbf{M}^{\dag}$ has only one element of $1$ in each row while all the other elements equal $0$, and the position of $1$ in the $\left(q+(b-1)(Q_{\mathrm{max}}+1)+(n-1)(Q_{\mathrm{max}}+1)(B_{\mathrm{max}}+1)\right)$-th row ($q\in\{1,\ldots,(Q_{\mathrm{max}}+1)\}, \ b\in\{1,\ldots,(B_{\mathrm{max}}+1)\}$, and $n\in\{1,\ldots,N\}$) corresponds to the position of the representative state $V(\widetilde{\mathbf{Q}}_{n}^{(q-1)},\widetilde{\mathbf{B}}_{n}^{(b-1)})$ (if $q\times b\neq 1$) or reference state $V(\mathbf{0},\mathbf{0})$ (if $q\times b=1$) in the global state vector $\mathbf{V}$. \par

Now the vector form of the equivalent Bellman post-decision equation \eqref{eq33} under all the representative states can be written as
   \begin{equation}
   \label{eq36}
      \theta\mathbf{e}+\widetilde{\mathbf{V}}_{\infty}(\boldsymbol{\eta})=\mathbf{M}^{\dag}\mathbf{T}\left(\boldsymbol{\eta},\mathbf{M}\widetilde{\mathbf{V}}_{\infty}(\boldsymbol{\eta})\right).
      \end{equation}
\noindent where $\mathbf{e}$ is a $(Q_{\mathrm{max}}+1)(B_{\mathrm{max}}+1)N\times 1$ vector with all elements equal to $1$. The mapping $\mathbf{T}$ is defined as
   \begin{displaymath}
   \mathbf{T}(\boldsymbol{\eta},\mathbf{V})=\min_{\Omega}\left[\mathbf{g}(\boldsymbol{\eta},\Omega)+\mathbf{P}(\Omega)\mathbf{V}\right]
   \end{displaymath}
\noindent where $\mathbf{g}(\boldsymbol{\eta},\Omega)$ is the vector form of function $g(\mathbf{Q},\mathbf{B},\Omega(\mathbf{Q},\mathbf{B}))$ defined in \eqref{eq17}, and $\mathbf{P}(\Omega)$ is the matrix form of transition probability $\mathrm{Pr.}[\widetilde{\mathbf{Q}}'|\mathbf{Q},\mathbf{B},\Omega(\mathbf{Q},\mathbf{B})\mathrm{Pr.}[\widetilde{\mathbf{B}}'|\mathbf{B},\Omega(\mathbf{Q},\mathbf{B})]$ defined in \eqref{eq20}.

\newtheorem{lemma3}[lemma]{Lemma}
\begin{lemma3}[Convergence of Per-node Value Function learning]
Denote
   \begin{displaymath}
   \mathbf{A}_{t-1}=(1-\varepsilon_{t-1}^{v})\mathbf{I}+\mathbf{M}^{\dag}\mathbf{P}(\Omega_{t})\mathbf{M}\varepsilon_{t-1}^{v}
   \end{displaymath}
   \begin{displaymath}
   \mathbf{B}_{t-1}=(1-\varepsilon_{t-1}^{v})\mathbf{I}+\mathbf{M}^{\dag}\mathbf{P}(\Omega_{t-1})\mathbf{M}\varepsilon_{t-1}^{v}
   \end{displaymath}
\noindent where $ \Omega_{t} $ is the unichain control policy at slot $t$, $\mathbf{P}(\Omega_{t})$ is the transition matrix under the unichain system control policy, and $ \mathbf{I} $ is the identity matrix. If for the entire sequence of control policies $ \{\Omega_{t}\}$ there exists $ \delta_{\beta}>0 $ and some positive integer $\beta$ such that
   \begin{displaymath}
   [\mathbf{A}_{\beta}\cdot\cdot\cdot\mathbf{A}_{1}]_{(z,I)}\geq\delta_{\beta}
   \end{displaymath}
   \begin{equation}
   \label{reference}
   [\mathbf{B}_{\beta}\cdot\cdot\cdot\mathbf{B}_{1}]_{(z,I)}\geq\delta_{\beta},\forall k
   \end{equation}
\noindent where $ [\cdot]_{(z,I)} $ denotes the element in the $z$-th row and the $I$-th column(where $I$ corresponds to the reference state that all queues and batteries are empty $(\mathbf{Q}_{I}=\mathbf{0},\mathbf{B}_{I}=\mathbf{0})$), and $ \delta_{t}=\mathcal{O}(\varepsilon_{t}^{v}) $. Then the following statements are true.
The update of the per-node value function vector will converge almost surely for any given initial parameter vector $\widetilde{\mathbf{V}}_{0}$ and LM vector $\boldsymbol{\eta}$, i.e.,
      \begin{displaymath}
      \lim_{t\rightarrow\infty}\widetilde{\mathbf{V}}_{t}(\boldsymbol{\eta})=\widetilde{\mathbf{V}}_{\infty}(\boldsymbol{\eta}).
      \end{displaymath}
\noindent The steady-state per-node value function vector $\widetilde{\mathbf{V}}_{\infty}$ satisfies \eqref{eq33}.\par

The proof of Lemma 3 is given in Appendix D.
\end{lemma3}

\newtheorem{remark4}[remark]{Remark}
\begin{remark4}[Interpretation of the Conditions in Lemma 2]
Note that $\mathbf{A}_{t}$ and $\mathbf{B}_{t}$ are related to an equivalent transition matrix of the underlying Markov chain. Equation \eqref{reference} simply means that the reference state $(\mathbf{0},\mathbf{0})$ is accessible from all the reduced states after some finite number of transition steps. This is a very mild condition and is satisfied in most of the cases we are interested.
\end{remark4}

\newtheorem{lemma4}[lemma]{Lemma}
\begin{lemma4}[Convergence of LMs Update]
The iteration on the vector of LMs $\boldsymbol{\eta}$ converges almost surely to the set of maxima of $G(\boldsymbol{\eta})$. Suppose the LMs converge to $\boldsymbol{\eta}^{*}$, then $\boldsymbol{\eta}^{*}$ satisfies the delay constraints in Problem 1.\par

The proof of Lemma 4 is given in Appendix E.
\end{lemma4}

\subsection{Asymptotic Performance Analysis}
Although Lemma 3 asserts that for any given LMs $\boldsymbol{\eta}$, the iteration on per-node value function will converge, there is still approximation error between the converged parameter vector $\widetilde{\mathbf{V}}^{\infty}(\boldsymbol{\eta})$ and original system value function $\mathbf{V}^{\infty}(\boldsymbol{\eta})$. In the following theorem, we provide a bound on the approximation error $\parallel \mathbf{M}\widetilde{\mathbf{V}}^{\infty}(\boldsymbol{\eta})-\mathbf{V}^{\infty}(\boldsymbol{\eta})\parallel$.

\newtheorem{Theorem}{Theorem}
\begin{Theorem}[Bound on Approximation Error]Let $\widetilde{\mathbf{V}}^{\infty}$ and $\mathbf{V}^{\infty}$ be the converged parameter vector and system value function vector, respectively, for any given LM $\boldsymbol{\eta}$, $\mathbf{X}^{*}=\arg\min\parallel\mathbf{M}\mathbf{X}-\mathbf{V}^{\infty}\parallel=(\mathbf{M}'\mathbf{M})^{-1}\mathbf{M}'\mathbf{V}^{\infty}$, $\mathbf{T}^{\dagger}(\mathbf{V})=\mathbf{T}(\mathbf{V})-T_{0}(\mathbf{V})\mathbf{e}$, $\widetilde{\mathbf{T}}(\mathbf{X})=\mathbf{M}^{\dagger}\mathbf{T}^{\dagger}(\mathbf{M}\mathbf{X})$, and $\widetilde{\mathbf{T}}^{(n)}(\mathbf{X})=\underbrace{\widetilde{\mathbf{T}}\circ\widetilde{\mathbf{T}}\circ\dots\circ\widetilde{\mathbf{T}}}_{n}$, where $T_{0}(\mathbf{V})$ denotes the element in the vector $\mathbf{T}(\mathbf{V})$ which corresponds to the reference system state that all queues are empty, the approximation error $\parallel \mathbf{M}\widetilde{\mathbf{V}}^{\infty}-\mathbf{V}^{\infty}\parallel$ is lower-bounded by
\begin{displaymath}
\parallel \mathbf{M}\widetilde{\mathbf{V}}^{\infty}-\mathbf{V}^{\infty}\parallel\geq\parallel \mathbf{M}\mathbf{X}^{*}-\mathbf{V}^{\infty}\parallel,
\end{displaymath}
\noindent and upper-bounded by
\begin{align*}
\parallel \mathbf{M}\widetilde{\mathbf{V}}^{\infty}-\mathbf{V}^{\infty}\parallel&\leq\frac{a(c^{n}+1)}{1-\beta}\parallel \mathbf{X}^{*}-\mathbf{M}^{\dagger}\mathbf{V}^{\infty}\parallel+\parallel\mathbf{M}\mathbf{X}^{*}-\mathbf{V}^{\infty}\parallel \\
&=\left(\mathbf{M}\mathbf{X}^{*}-\mathbf{V}^{\infty}\right)^{T}\left(\frac{a(c^{n}+1)}{1-\beta}\mathbf{M}^{\dagger'}\mathbf{M}^{\dagger}+\mathbf{I}\right)\left(\mathbf{M}\mathbf{X}^{*}-\mathbf{V}^{\infty}\right),
\end{align*}
\noindent where $a=\parallel\mathbf{M}\parallel$ denotes the 2-norm of the matrix $\mathbf{M}$, which satisfies $0\leq a<\infty$ due to the mathematical property of 2-norm, integer $n$ and $0<\beta<1$ should satisfy
\begin{displaymath}
\parallel \widetilde{\mathbf{T}}^{(n)}(\widetilde{\mathbf{V}}^{\infty})-\widetilde{\mathbf{T}}^{(n)}(\mathbf{X}^{*})\parallel\leq\beta\parallel \widetilde{\mathbf{V}}^{\infty}-\mathbf{X}^{*}\parallel,
\end{displaymath}
\noindent and $c$ should satisfy
\begin{displaymath}
\parallel \widetilde{\mathbf{T}}^{(m)}(\mathbf{X}^{*})-\mathbf{M}^{\dagger}\mathbf{V}^{\infty}\parallel\leq c\parallel\widetilde{\mathbf{T}}^{(m-1)}(\mathbf{X}^{*})-\mathbf{M}^{\dagger}\mathbf{V}^{\infty}\parallel, m=1,2,\dots,n.
\end{displaymath}

The proof of Theorem 1 is given in Appendix F.
\end{Theorem}

\newtheorem{remark5}[remark]{Remark}
\begin{remark5}[Discussion on Approximation Error]
From the above theorem, we observe that
\begin{itemize}
\item[1)] Due to the convergence property of per-node value iteration, we have $\lim\limits_{n\rightarrow\infty}\widetilde{\mathbf{T}}^{(n)}(\mathbf{X}^{*})=\widetilde{\mathbf{V}}^{\infty}$. Hence, there always exists at least one pair of $(n,\beta)$ such that $\parallel \widetilde{\mathbf{T}}^{(n)}(\widetilde{\mathbf{V}}^{\infty})-\widetilde{\mathbf{T}}^{(n)}(\mathbf{X}^{*})\parallel\leq\beta\parallel \widetilde{\mathbf{V}}^{\infty}-\mathbf{X}^{*}\parallel$. Intuitively, $n$ and $\beta$ measure the convergence speed of mapping $\widetilde{\mathbf{T}}$: smaller $n$ or smaller $\beta$ means higher convergence speed.
\item[2)] Note that $\parallel \widetilde{\mathbf{T}}^{(m)}(\mathbf{X}^{*})-\mathbf{M}^{\dagger}\mathbf{V}^{\infty}\parallel=\parallel \mathbf{M}^{\dagger}\mathbf{T}^{\dagger}(\mathbf{M}\widetilde{\mathbf{T}}^{(m-1)}(\mathbf{X}^{*}))-\mathbf{M}^{\dagger}\mathbf{T}^{\dagger}(\mathbf{V}^{\infty})\parallel$ and $\mathbf{T}^{\dagger}$ is a contraction mapping, there always exists a sufficient large $c$ such that $\parallel \widetilde{\mathbf{T}}^{(m)}(\mathbf{X}^{*})-\mathbf{M}^{\dagger}\mathbf{V}^{\infty}\parallel\leq c\parallel\widetilde{\mathbf{T}}^{(m-1)}(\mathbf{X}^{*})-\mathbf{M}^{\dagger}\mathbf{V}^{\infty}\parallel$ is satisfied. Intuitively, $c$ measures the ``contraction ratio" of the mapping $\mathbf{T}^{\dagger}$ on the reference states $\mathbf{S}_{I}$: the smaller $c$, the larger the contraction ratio.
\item[3)] From the above discussion, if the mapping $\widetilde{\mathbf{T}}$ has good convergence speed and the mapping $\mathbf{T}^{\dagger}$ has large contraction ratio on the representative states, the upper-bound of approximation error will become small (due to small $\frac{a(c^{n}+1)}{1-\beta}$); however, the approximation error will never be smaller than $\parallel\mathbf{M}\mathbf{X}^{*}-\mathbf{V}^{\infty}\parallel$, which is because of the fundamental limitation on the vector dimension.
\end{itemize}
\end{remark5}

\section{Experimental Results}
In this section, we conduct experiments to evaluate the performance of our proposed AMDP+OSL algorithm, and compare it with the offline value iteration (OVI) algorithm and other proposed algorithms in literature. In the simulations, we consider Poisson packet arrival with mean arrival rate $\lambda^{\mathrm{A}}$ at every EHS node with fixed packet size of 100K bits \cite{TWC:Sharma}. Moreover, we consider that the fundamental energy unit is 1J and the EHS node either harvests $2\lambda^{\mathrm{E}}$ units of energy or does not harvest any at each time slot with equal probability \cite{Sudevalayam:CST}, i.e., $E_{n,t}\in \{0,2\lambda^{\mathrm{E}}\}$ and $\mathbf{E}[E_{n}]=\lambda^{\mathrm{E}}$ for all $n\in\{1,\ldots,N\}$. This process imitates the solar cycles for a solar battery during the day and night. (We have also performed simulations when the amount of energy arrived per time slot has truncated Poisson, Exponential, Erlang or Hyperexponential distributions and found that conclusions drawn in this section continue to hold.) The wireless channel bandwidth is $W=0.3\mathrm{MHz}$, and the noise power spectral density is $N_{0}=10^{-16} \mathrm{W/Hz}$. The channel state can be ``G=Good", ``N=Normal", or ``B=Bad", corresponding to the channel gain ``$6\times 10^{-13}$", ``$4\times 10^{-13}$", or ``$2\times 10^{-13}$", respectively. The fast fading of the wireless channel is captured by the three-state Markov chain with the transition matrix given by \cite{TVT:Mao}
\begin{displaymath}
P_{H}=\left[
\begin{array}{ccc}
    P_{BB} & P_{BN} & P_{BG}  \\
    P_{NB} & P_{NN} & P_{NG}  \\
    P_{GB} & P_{GN} & P_{GG}  \\
\end{array}
\right]=\left[
\begin{array}{ccc}
    0.3 & 0.7 & 0  \\
    0.25 & 0.5 & 0.25  \\
    0 & 0.7 & 0.3  \\
\end{array}
\right],
\end{displaymath}
\noindent where $P_{XZ}$ represents the probability of the channel state evolving from state $X$ to state $Z$, with $X,Z\in\{B,N,G\}$. We choose target BER $\epsilon_{n}=10^{-3}$ and $\xi_{n}\approx 0.283$ \cite{TVT:Mao}. The data buffer and the rechargeable battery capacity are $Q_{\max}=5$ and $B_{\max}=10$, respectively. The data sensing efficiency parameter $\gamma$ varies from 0.6 to 1.4 \cite{TVT:Mao}. We assign equal weight of packet loss rate on each EHS node. \par

In order to analyze performance in different aspects and evaluate different algorithms, we divide this section into two parts. Firstly, we consider the single node scenario, and concentrate on the comparison of different energy allocation algorithms. Then, we consider the multiple node scenario and analyze the scheduling and energy allocation algorithms jointly.

\subsection{Single Node Network}
We compare our proposed AMDP+OSL algorithm with the OVI algorithm based on the post-decision equivalent Bellman's equation \eqref{eq20}, the OEA algorithm \cite{TVT:Mao}, the MWF algorithm \cite{TWC:Sharma} and Energy Adaptive Water-Filling (EAWF) algorithm \cite{JSAC:Ozel}. The OEA algorithm finds an optimal sensing and transmit energy allocation policy which maximizes the expected amount of transmitted data under the assumption that there is always data to sense. The MWF algorithm is a modified water-filling policy which minimizes the mean delay, where the sensing energy is considered as a stationary, ergodic sequence. The EAWF algorithm maximizes the number of bits transmitted by the deadline without considering the sensing energy, which is an event-based online policy reacting to a change in fading level or an energy arrival. Since \eqref{eq20} is equivalent to the original Bellman's equation \eqref{eq16} for the optimal MDP problem, we will refer to it as OMDP+OVI algorithm in the following discussion. The performance of the OMDP+OVI algorithm is the optimal value and provides an upper bound for all the algorithms. The OEA algorithm also uses offline value iteration to derive the control policy, so its computation complexity is the same with that of the OMDP+OVI algorithm and both algorithms cannot be extended to the multi-node scenario due to curse of dimensionality. The average delay constraint is $D_{\max}=3$ time slots.\par

\begin{figure}
\centering
\includegraphics[width=0.52\textwidth]{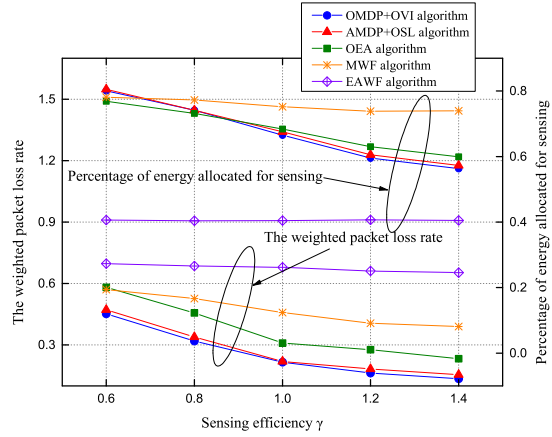}
\vspace{-2em}
\caption{The weighted packet loss rate and percentage of energy allocated for sensing versus sensing efficiency $\gamma$ for single node network ($\lambda^{\mathrm{E}}=1.2$ and $\lambda^{\mathrm{A}}=1$).}
\label{sensing}
\end{figure}

Fig.\ref{sensing} shows the weighted packet loss rate and percentage of energy allocated for sensing versus sensing efficiency $\gamma$. It can be observed that as the sensing efficiency $\gamma$ increases, the percentages of allocated energy for sensing decrease, and lower weighted packet loss rates are achieved for all four algorithms. The allocated sensing energy and the performance of our proposed AMDP+OSL algorithm are quite close to the optimal values achieved by the OMDP+OVI algorithm. On the other hand, both the OEA algorithm and MWF algorithm do not respond to the varying sensing efficiency $\gamma$ as accurately as the AMDP+OSL and OMDP+OVI algorithms in energy allocation. The sensing energy allocation of the MWF algorithm remains approximately the same irrespective of the varying sensing efficiency, because the sensing energy is considered as a stationary, ergodic sequence. On the other hand, the sensing energy allocation of the OEA algorithm is smaller than the optimal value when the sensing efficiency $\gamma$ is small, while larger than the optimal value when the sensing efficiency $\gamma$ is large, so its performance is worse than our proposed AMDP+OSL algorithm. The percentage of energy allocated for sensing by the EAWF algorithm stays extremely low, which leads to the highest packet loss rate, since the energy allocation of EAWF algorithm just optimizes the transmission energy for maximum throughput without considering the energy for sensing.\par

\begin{figure}
\centering
\includegraphics[width=0.52\textwidth]{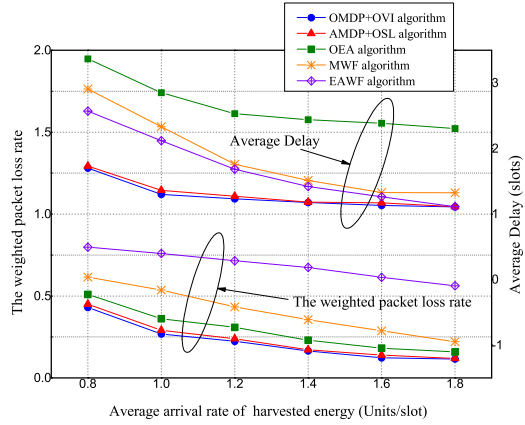}
\vspace{-2em}
\caption{The weighted packet loss rate and average delay versus the average arrival rate of harvested energy $\lambda^{\mathrm{E}}$ for single node network ($\gamma=1$ and $\lambda^{\mathrm{A}}=1$).}
\label{Earrive1}
\end{figure}
\begin{figure}
\centering
\includegraphics[width=0.52\textwidth]{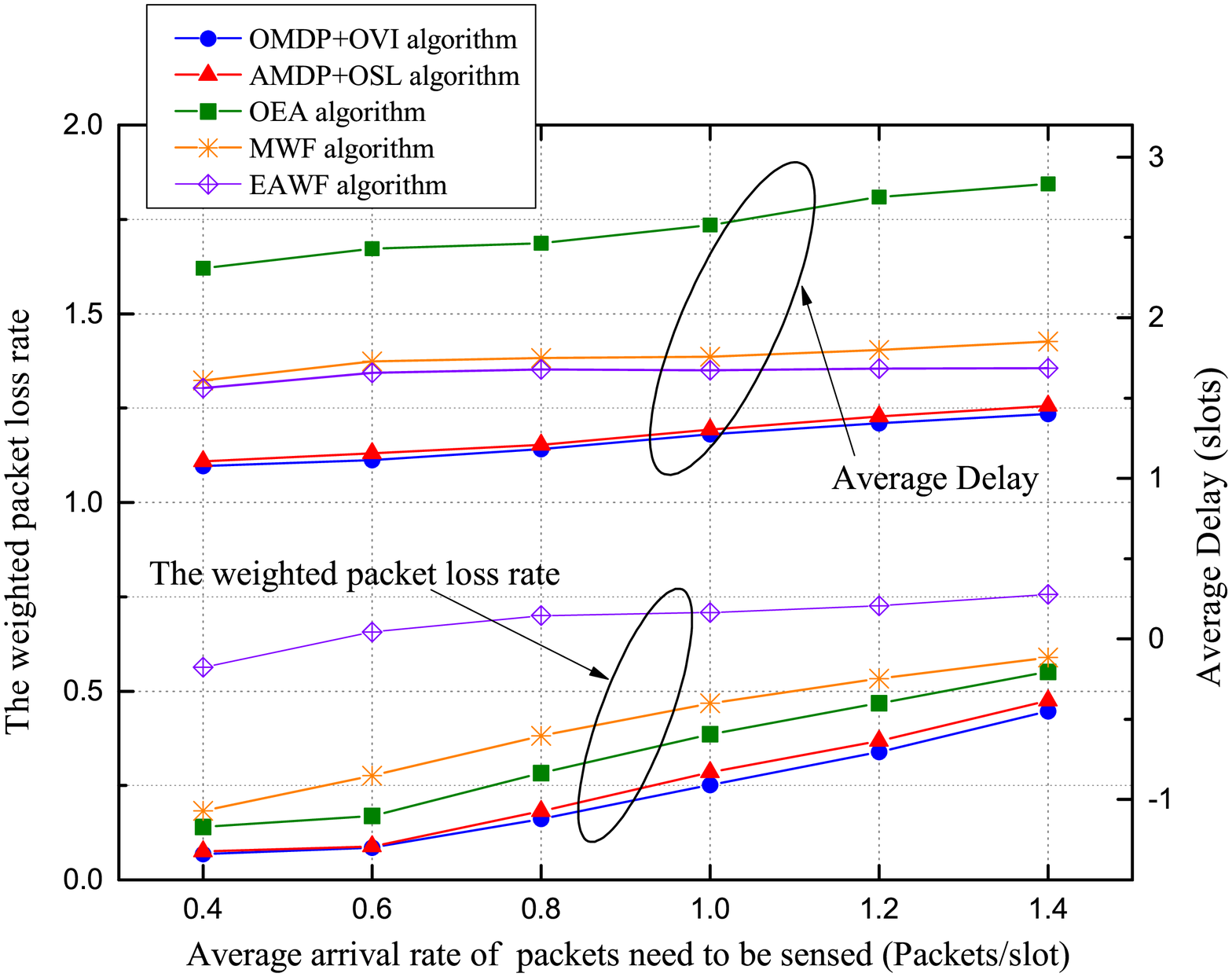}
\vspace{-2em}
\caption{The weighted packet loss rate and and average delay versus the average arrival rate of packets $\lambda^{\mathrm{A}}$ for single node network ($\gamma=1$ and $\lambda^{\mathrm{E}}=1.2$).}
\label{Aarrive1}
\end{figure}

Fig.\ref{Earrive1} and Fig.\ref{Aarrive1} show the weighted packet loss rate and the average delay versus the average arrival rates of harvested energy and packets, respectively. The packet loss rate and average delay of all the simulated algorithms decrease with increasing harvested energy arrival rate and decreasing packet arrival rate as expected. It can be observed that our proposed AMDP+OSL algorithm achieves smaller packet loss rate than the OEA and MWF algorithms while satisfying the delay constraint. The OEA algorithm results in much larger average delay than the other algorithms, since it does not take the packet arrival rate into consideration, and may waste energy for sensing and result in less energy for data transmission. Moreover, the complexity and signaling overhead of the OEA algorithm is the same with that of the OMDP+OVI algorithm, and is much larger than our proposed algorithm. The MWF algorithm has the second highest packet loss rate, because the sensing energy is set to be a stationary, ergodic sequence, which does not change with the varying packet and energy arrival rates. The EAWF algorithm achieves lower average delay than OEA and MWF algorithms, since the packets sensed successfully are limited and the energy for transmission is abundant. However, the EAWF algorithm gets higher average delay than our proposed algorithm, since it assumes the infinite backlog traffic model and doesn't consider the packet arrival rate. Compared with the OMDP+OVI algorithm, our proposed algorithm results in a little performance loss due to the linear value function approximation and Taylor's expansion in obtaining the optimal action. According to the statistics, the packet loss rate of AMDP+OSL algorithm is just 3\% higher than that of OMDP+OVI algorithm. Therefore, the AMDP+OSL algorithm is an effective method to reduce the complexity and signaling overhead, while achieving a near optimal performance. \par

%
%
%

\subsection{Multiple Node Network}
We compare our proposed algorithm with the OVI algorithm, the UROP scheduling algorithm \cite{Gul:WCNC} and Greedy scheduling algorithm \cite{Sharma:Allerton}. The UROP algorithm schedules the nodes according to their energy under the assumption that each node always has data to transmit and the energy storage buffer is infinite. The Greedy algorithm is an opportunistic scheduler that minimizes the weighted sum of mean delays where the energy storage buffer is assumed to be infinite. We consider that MWF energy allocation algorithm \cite{TWC:Sharma} is used with the latter two scheduling algorithms. Since the curse of dimensionality problem forbids the implementation of OVI algorithm based on \eqref{eq20}, we apply the OVI algorithm to the Bellman's equation \eqref{eq33} for representative states after applying the approximate MDP algorithm and refer to it as AMDP+OVI algorithm. The average delay constraint is set to $D_{\max}=8$ time slots.\par

\begin{figure}
\centering
\includegraphics[width=0.52\textwidth]{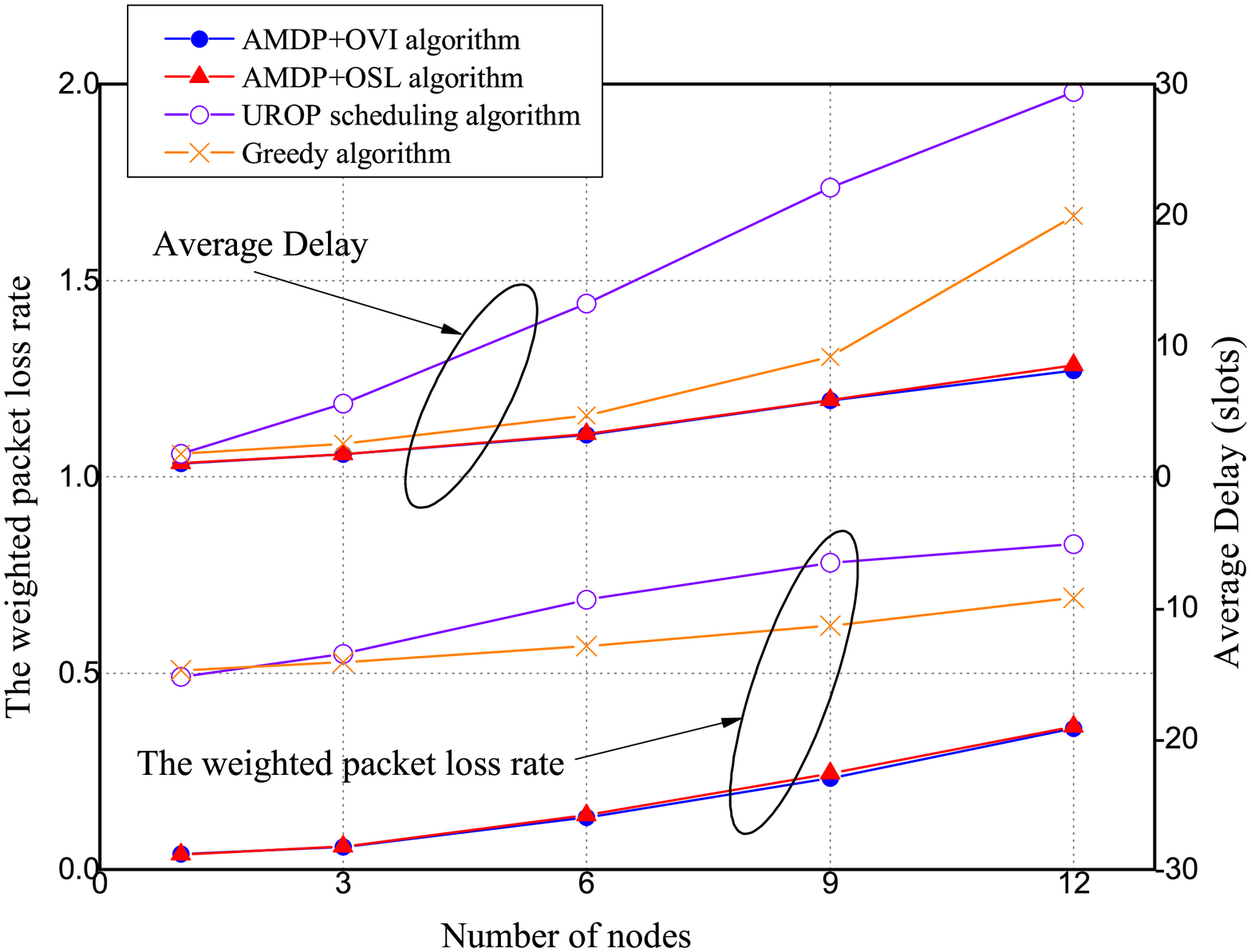}
\vspace{-2em}
\caption{The weighted packet loss rate and the average delay versus the number of EHS nodes $N$ for multiple node network ($\lambda^{\mathrm{E}}=1.2$ and $\lambda^{\mathrm{A}}=0.8$).}
\label{node}
\end{figure}

Fig.\ref{node} shows the weighted packet loss rate and the average delay versus the number of EHS nodes. It is obvious that our proposed AMDP+OSL algorithm has a significant gain in packet loss rate over the two other reference algorithms while satisfying the delay constraint. This is in part due to the near optimal energy allocation achieved by AMDP+OSL algorithm as discussed in the single node scenario, and also due to its near optimal scheduling policy. The Greedy algorithm considers the CSI and QSI but not the BSI, which leads to higher packet loss rate and average delay compared to AMDP+OSL algorithm. The scheduling process of UROP algorithm is random without considering the system state, which leads to the highest packet loss rate and average delay. The performance of AMDP+OVI algorithm and AMDP+OSL algorithm are nearly the same, which shows that the OSL algorithm can converge to the optimal value of OVI algorithm without demanding the probability distributions of the arrival processes and channel characteristics. \par

\begin{figure}
\centering
\includegraphics[width=0.52\textwidth]{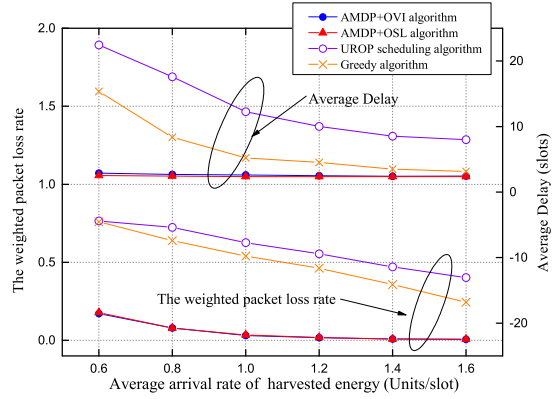}
\vspace{-2em}
\caption{The weighted packet loss rate and average delay versus the average arrival rate of harvested energy $\lambda^{\mathrm{E}}$ for multiple node network ($N=5$ and $\lambda^{\mathrm{A}}=0.5$).}
\label{Earrive2}
\end{figure}
\begin{figure}
\centering
\includegraphics[width=0.52\textwidth]{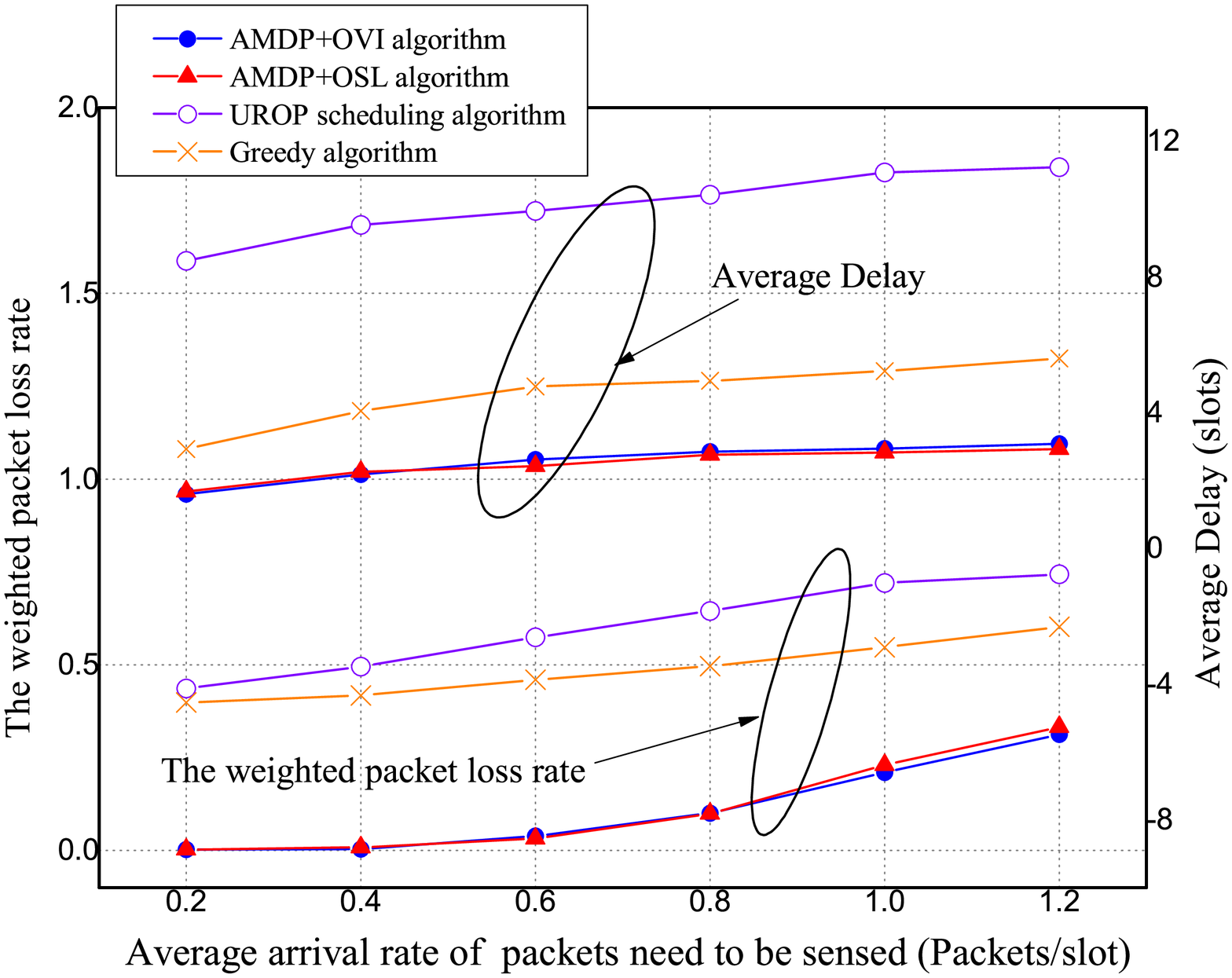}
\vspace{-2em}
\caption{The weighted packet loss rate and and average delay versus the average arrival rate of packets $\lambda^{\mathrm{A}}$ ($N=5$ and $\lambda^{\mathrm{E}}=1.2$).}
\label{Aarrive2}
\end{figure}

Fig.\ref{Earrive2} and Fig.\ref{Aarrive2} show the weighted packet loss rate and the average delay versus the average arrival rates of harvested energy and packets, respectively. It can be observed that our proposed AMDP+OSL algorithm can always achieve the lowest packet loss rate while satisfying the delay constraint. In addition, when the average arrival rate of harvested energy or packets is low, the packet loss rate of Greedy algorithm and UROP algorithm are nearly the same. This is because the packets sensed successfully are limited, and the packet loss rate mainly comes from the packet drop during the sensing process. On the other hand, when the average arrival rate of harvested energy or packets increases, the successfully sensed packets increase and are backlogged in the buffer. In this case, the gap of packet loss rate between the Greedy algorithm and UROP algorithm becomes large due to the larger data buffer overflow of the UROP scheduling algorithm. \par

\section{Conclusions and Future Work}
In this paper, we have proposed an energy allocation and scheduling algorithm to minimize the weighted packet loss rate with delay constraint for IWSN. We have shown that the reliability optimal control problem can be formulated as a CMDP. To reduce the complexity and facilitate the distributed implementation, we have utilized approximate MDP and online stochastic learning to obtain a distributed energy allocation algorithm with multi-level water-filling structure based on local system state and an auction mechanism to determine the scheduled EHS node per time slot. The simulation results have shown that the reliability performance of our proposed algorithm under delay constraint is very close to that achieved by the offline value iteration algorithm, and is significantly better than various baselines. The comparison results indicate that in order to achieve optimal reliability in EH-IWSN, the energy allocation for both the sensing and transmission processes need to be considered. Moreover, the energy allocation and scheduling algorithms should be jointly optimized based on the CSI, QSI, and BSI. The proposed algorithm can be used in many applications of the EH-IWSN, where the monitoring system
has to provide accurate and real-time information regarding the monitored process, such as river flood detection, fence surveillance, and equipment condition monitoring of pipelines and machinery, etc.\par

Although our focus is on single-hop IWSN with a star topology in this paper, where the FC is the sink node, the proposed algorithm can be applied to multi-hop IWSN with a cluster-tree topology, where a cluster head is responsible for gathering data from the cluster's EHS nodes and forwarding it to the sink. On the other hand, it is a non-trivial problem to extend the proposed algorithm to multi-hop networks with mesh topology due to the complex coupled queue dynamics. Moreover, it is of interest to study the impact of imperfect or delayed CSI at the FC. Our proposed algorithm can be extended to include leakage during storage process if the amount of leakage is assumed to be a fixed value.\par

\appendix

\subsection{Proof of Lemma 1}
\begin{displaymath}
\begin{split}
&\theta+V(\mathbf{H},\mathbf{Q},\mathbf{B}) \ \forall \mathbf{H}\in\mathcal{H}, \ \mathbf{Q}\in\mathcal{Q},\mathbf{B}\in\mathcal{B}\\
=&\min_{\Omega(\mathbf{S})} \Big\{g(\mathbf{H},\mathbf{Q},\mathbf{B},\Omega(\mathbf{H},\mathbf{Q},\mathbf{B}))
+\sum_{\mathbf{H}',\mathbf{Q}',\mathbf{B}'}\mathrm{Pr.}[\mathbf{H}',\mathbf{Q}',\mathbf{B}'|
\mathbf{H},\mathbf{Q},\mathbf{B},\Omega(\mathbf{H},\mathbf{Q},\mathbf{B})]V(\mathbf{H}',\mathbf{Q}',\mathbf{B}')\Big\} \\
\overset{(a)}=&\min_{\Omega(\mathbf{S})}\Big\{
g(\mathbf{H},\mathbf{Q},\mathbf{B},\Omega(\mathbf{H},\mathbf{Q},\mathbf{B}))+\sum_{\mathbf{Q}',\mathbf{B}'}\mathrm{Pr.}[\mathbf{Q}'|\mathbf{H},\mathbf{Q},
\Omega(\mathbf{H},\mathbf{Q},\mathbf{B})]
\\
&\mathrm{Pr.}[\mathbf{B}'|\mathbf{B},
\Omega(\mathbf{H},\mathbf{Q},\mathbf{B})]\big(\sum_{\mathbf{H}'}\mathrm{Pr.}(\mathbf{H}'|\mathbf{H})V(\mathbf{H}',\mathbf{Q}',\mathbf{B}')\big)\Big\}  \\
\overset{(b)}=&\min_{\Omega(\mathbf{S})}\Big\{g(\mathbf{H},\mathbf{Q},\mathbf{B},\Omega(\mathbf{H},\mathbf{Q},\mathbf{B})) \\
&+\sum_{\mathbf{Q}',\mathbf{B}'}\mathrm{Pr.}[\mathbf{Q}'|\mathbf{H},\mathbf{Q},\Omega(\mathbf{H},\mathbf{Q},\mathbf{B})]
\mathrm{Pr.}[\mathbf{B}'|\mathbf{B},
\Omega(\mathbf{B})]\mathbf{E}_{\mathbf{H}'|\mathbf{H}}[V(\mathbf{H}',\mathbf{Q}',\mathbf{B}')|\mathbf{Q}',\mathbf{B}']\Big\}, \\
\end{split}
\end{displaymath}
\noindent where (a) is due to \eqref{eq4} by the independence between $(\mathbf{H},\mathbf{Q})$ and $\mathbf{B}$ over time slots. \par

Taking the conditional expectation (conditioned on $(\mathbf{Q},\mathbf{B})$) on both sides of the equation above, we have
\begin{displaymath}
\begin{split}
&\theta+V(\mathbf{Q},\mathbf{B}) \ \forall \mathbf{Q}^{(l)}\in\mathcal{Q},\mathbf{B}^{(v)}\in\mathcal{B}, \\
=&\mathbf{E}_{\mathbf{H}}\Big[\min_{\Omega(\mathbf{S})}\Big\{g(\mathbf{H},\mathbf{Q},\mathbf{B},\Omega(\mathbf{H},\mathbf{Q},\mathbf{B}))
\\
&+\sum_{\mathbf{Q}',\mathbf{B}'}\mathrm{Pr.}[\mathbf{Q}'|\mathbf{H},\mathbf{Q},
\Omega(\mathbf{H},\mathbf{Q},\mathbf{B})]\mathrm{Pr.}[\mathbf{B}'|\mathbf{B},
\Omega(\mathbf{H},\mathbf{Q},\mathbf{B})]\mathbf{E}_{\mathbf{H}'|\mathbf{H}}[V(\mathbf{H}',\mathbf{Q}',\mathbf{B}')|\mathbf{Q}',\mathbf{B}'])\Big\}\Big]\\
\overset{(c)}=&\min_{\Omega(\mathbf{Q},\mathbf{B})}\Big\{g(\mathbf{Q},\mathbf{B},\Omega(\mathbf{Q},\mathbf{B}))
+\sum_{\mathbf{Q}',\mathbf{B}'}\mathrm{Pr.}[\mathbf{Q}'|\mathbf{Q},\Omega(\mathbf{Q},\mathbf{B})]\mathrm{Pr.}[\mathbf{B}'|\mathbf{B},\Omega(\mathbf{Q},\mathbf{B})]V(\mathbf{Q}',\mathbf{B}')\Big\},
\\
\end{split}
\end{displaymath}
\noindent where $(c)$ is due to the definition of $g(\mathbf{Q},\mathbf{B},\Omega(\mathbf{Q},\mathbf{B}))$, $\mathrm{Pr.}[\mathbf{Q}'|\mathbf{Q},\Omega(\mathbf{Q},\mathbf{B})]$, $\mathrm{Pr.}[\mathbf{B}'|\mathbf{B},\Omega(\mathbf{Q},\mathbf{B})]$ and $V(\mathbf{Q}',\mathbf{B}')$ in Section V.B.\par

\subsection{Proof of Lemma 2}

\begin{displaymath}
\begin{split}
&\theta+V(\mathbf{Q},\mathbf{B}) \ \forall \mathbf{Q}\in\mathcal{Q},\mathbf{B}\in\mathcal{B},\\
=&\min_{\Omega(\mathbf{Q},\mathbf{B})}\Big\{g(\mathbf{Q},\mathbf{B},\Omega(\mathbf{Q},\mathbf{B}))+\sum_{\mathbf{Q}',\mathbf{B}'}\mathrm{Pr.}[\mathbf{Q}'|\mathbf{Q},\mathbf{B},\Omega(\mathbf{Q},\mathbf{B})]\mathrm{Pr.}[\mathbf{B}'|\mathbf{B},\Omega(\mathbf{Q},\mathbf{B})]V(\mathbf{Q}',\mathbf{B}')\Big\} \\
=&\min_{\Omega(\mathbf{Q},\mathbf{B})}\Big\{g(\mathbf{Q},\mathbf{B},\Omega(\mathbf{Q},\mathbf{B}))\\
&+\sum_{\mathbf{Q}',\widetilde{\mathbf{Q}}'}\mathrm{Pr.}[\mathbf{Q}'|\widetilde{\mathbf{Q}}',\widetilde{\mathbf{B}}']\mathrm{Pr.}[\widetilde{\mathbf{Q}}'|\mathbf{Q},\Omega(\mathbf{Q},\mathbf{B})] \sum_{\mathbf{B}',\widetilde{\mathbf{B}}'}\mathrm{Pr.}[\mathbf{B}'|\widetilde{\mathbf{B}}']\mathrm{Pr.}[\widetilde{\mathbf{B}}'|\mathbf{B},\Omega(\mathbf{Q},\mathbf{B})]V(\mathbf{Q}',\mathbf{B}')\Big\} \\
=&\min_{\Omega(\mathbf{Q},\mathbf{B})}\Big\{g(\mathbf{Q},\mathbf{B},\Omega(\mathbf{Q},\mathbf{B}))\\
&+\sum_{\widetilde{\mathbf{Q}}',\widetilde{\mathbf{B}}'}\mathrm{Pr.}[\widetilde{\mathbf{Q}}'|\mathbf{Q},\Omega(\mathbf{Q},\mathbf{B})]\mathrm{Pr.}[\widetilde{\mathbf{B}}'|\mathbf{B},\Omega(\mathbf{Q},\mathbf{B})]\big(\sum_{\mathbf{Q}',\mathbf{B}'}\mathrm{Pr.}[\mathbf{Q}'|\widetilde{\mathbf{Q}}',\widetilde{\mathbf{B}}']\mathrm{Pr.}[\mathbf{B}'|\widetilde{\mathbf{B}}']V(\mathbf{Q}',\mathbf{B}')\big)\Big\} \\
\overset{(a)}=&\min_{\Omega(\mathbf{Q},\mathbf{B})}\Big\{g(\mathbf{Q},\mathbf{B},\Omega(\mathbf{Q},\mathbf{B}))+\sum_{\widetilde{\mathbf{Q}}',\widetilde{\mathbf{B}}'}\mathrm{Pr.}[\widetilde{\mathbf{Q}}'|\mathbf{Q},\Omega(\mathbf{Q},\mathbf{B})]\mathrm{Pr.}[\widetilde{\mathbf{B}}'|\mathbf{B},\Omega(\mathbf{Q},\mathbf{B})]V(\widetilde{\mathbf{Q}}',\widetilde{\mathbf{B}}')\Big\},
\end{split}
\end{displaymath}
\noindent where $(a)$ is due to the definition $V(\widetilde{\mathbf{Q}}',\widetilde{\mathbf{B}}')$ given in Section V.C. \par

Taking the conditional expectation (conditioned on $(\widetilde{\mathbf{Q}},\widetilde{\mathbf{B}})$) on both sides of the equation above, we have
\begin{displaymath}
\begin{split}
&\theta+V(\widetilde{\mathbf{Q}},\widetilde{\mathbf{B}}) \ \forall \widetilde{\mathbf{Q}}\in\mathcal{Q},\widetilde{\mathbf{B}}\in\mathcal{B},\\
=&\sum_{\mathbf{Q},\mathbf{B}}\mathrm{Pr.}[\mathbf{Q}|\widetilde{\mathbf{Q}},\widetilde{\mathbf{B}}]\mathrm{Pr.}[\mathbf{B}|\widetilde{\mathbf{B}}]\\&\min_{\Omega(\mathbf{Q},\mathbf{B})}\Big\{g(\mathbf{Q},\mathbf{B},\Omega(\mathbf{Q},\mathbf{B}))+\sum_{\widetilde{\mathbf{Q}}',\widetilde{\mathbf{B}}'}\mathrm{Pr.}[\widetilde{\mathbf{Q}}'|\mathbf{Q},\Omega(\mathbf{Q},\mathbf{B})]\mathrm{Pr.}[\widetilde{\mathbf{B}}'|\mathbf{B},\Omega(\mathbf{Q},\mathbf{B})]V(\widetilde{\mathbf{Q}}',\widetilde{\mathbf{B}}')\Big\}.
\end{split}
\end{displaymath}

\subsection{Proof of Algorithm 1}

In order to reduce the complexity of solving \eqref{eq25}, we first introduce the following lemma on the property of the optimal power allocation action under every system state.
\newtheorem{lemma8}[lemma]{Lemma}
\begin{lemma8}[Property of Optimal Policy]
The optimal policy $\Omega^{*}(\mathbf{S})=(x^{*},p^{\mathrm{T}*})$ satisfy $r\left(H_{n},p_{n}^{\mathrm{T}*}\right)\leq Q_{n}$ for any $n=1,\ldots,N$.
\end{lemma8}


With the constraint defined by Lemma 3, \eqref{eq26} becomes $Q_{n}(r(H_{n},p_{n}^{\mathrm{T}}))=Q_{n}-r(H_{n},p_{n}^{\mathrm{T}})$. Next, we expand $\widetilde{V}_{n}\left(Q_{n}\left(r(H_{n},p_{n}^{\mathrm{T}})\right),B_{n}\left(p_{n}^{\mathrm{T}}\right)\right)$ in \eqref{eq25} using Taylor expansion
\begin{equation}
\label{eq28}
\widetilde{V}_{n}\left(Q_{n}\left(r(H_{n},p_{n}^{\mathrm{T}})\right),B_{n}\left(p_{n}^{\mathrm{T}}\right)\right)=\widetilde{V}_{n}(Q_{n},B_{n})-r(H_{n},p_{n}^{\mathrm{T}})\big(\widetilde{V}_{n}^{(Q)}(Q_{n},B_{n})\big)'-p_{n}^{\mathrm{T}}\big(\widetilde{V}_{n}^{(B)}(Q_{n},B_{n})\big)'
\end{equation}
\noindent where
\begin{displaymath}
\big(\widetilde{V}_{n}^{(Q)}(Q_{n},B_{n})\big)'\approx \widetilde{V}_{n}(Q_{n}+1,B_{n})/2-\widetilde{V}_{n}(Q_{n}-1,B_{n})/2.
\end{displaymath}
\begin{displaymath}
\big(\widetilde{V}_{n}^{(B)}(Q_{n},B_{n})\big)'\approx \widetilde{V}_{n}(Q_{n},B_{n}+1)/2-\widetilde{V}_{n}(Q_{n},B_{n}-1)/2.
\end{displaymath}

Therefore, \eqref{eq25} is equivalent to
\begin{equation}
\label{eq29}
\Omega^{*}(\mathbf{S})
=\arg\max_{\Omega(\mathbf{S})}\sum_{n=1}^{N}bid_{n}
\end{equation}
\begin{displaymath}
\begin{split}
\textrm{s.t. } &0\leq\ p_{n}^{\mathrm{T}}\leq \min[B_{n},\frac{N_{0}\tau W (2^{\frac{Q_{n}K}{\tau W}}-1)}{\xi_{n}H_{n}}], \forall n=1,\ldots,N, \\
&\sum_{n=1}^{N}\mathbf{I}(p_{n}^{\mathrm{T}}>0)\leq 1,
\end{split}
\end{displaymath}
\noindent where $bid_{n}$ is given in \eqref{eq30}. The constraints are due to the definition of action space in Section IV and the property of optimal policy in Lemma 3.\par

\subsection{Proof of Lemma 3}
Since each representative state is updated comparably often in the asynchronous learning algorithm, quoting the conclusion in \cite{SIAM:Asynchronous}, the convergence property of the asynchronous update and the synchronous update is the same. Therefore, we consider the convergence of the related synchronous version for simplicity in this proof. Similar to \cite{SIAM:Borkar}, it is easy to see that the per-node value function vector $\widetilde{\mathbf{V}}_{t}$ is bounded almost surely during the iterations of the algorithm. In the following, we first introduce and prove the following lemma on the convergence of learning noise.\par

\newtheorem{lemma6}[lemma]{Lemma}
\begin{lemma6}Define
\begin{displaymath}
\mathbf{q}_{t}=\mathbf{M}^{\dag}[\mathbf{g}(\Omega_{t})+\mathbf{P}(\Omega_{t})\mathbf{M}\widetilde{\mathbf{V}}_{t}-\mathbf{M}\widetilde{\mathbf{V}}_{t}-\mathbf{T}_{0}(\mathbf{M}\widetilde{\mathbf{V}}_{t})\mathbf{e}],
\end{displaymath}
\noindent where $ \mathbf{T}_{0}(\mathbf{V})=\min_{\Omega}\left[\mathbf{g}_{I}(\Omega)+\mathbf{P}_{I}(\Omega)\mathbf{V}\right] $ denotes the mapping on the state $(\mathbf{Q}_{I},\mathbf{B}_{I})$, where $ \mathbf{g}_{I}(\Omega) $ is the vector form of function $ g(\mathbf{Q}_{I},\mathbf{B}_{I},\Omega(\mathbf{Q}_{I},\mathbf{B}_{I})) $, $ \mathbf{P}_{I}(\Omega) $ is the matrix form of transition probability $\mathrm{Pr.}[\widetilde{\mathbf{Q}}'|\mathbf{Q}_{I},\Omega(\mathbf{Q}_{I},\mathbf{B}_{I})]\mathrm{Pr.}[\widetilde{\mathbf{B}}'|\mathbf{B}_{I},\Omega(\mathbf{Q}_{I},\mathbf{B}_{I})]$. When the number of iterations $ t\geq j\rightarrow\infty $, the procedure of update can be written as follows with probability 1:
\begin{displaymath}
\widetilde{\mathbf{V}}_{t+1}=\widetilde{\mathbf{V}}_{j}+\sum_{i=j}^{t}\varepsilon_{i}^{v}\mathbf{q}_{i}.
\end{displaymath}
\end{lemma6}

\begin{IEEEproof}
The synchronous update of per-node value function vector can be written in the following vector form:
\begin{displaymath}
\widetilde{\mathbf{V}}_{t+1}=\widetilde{\mathbf{V}}_{t}+\varepsilon_{i}^{v}\mathbf{M}^{\dag}[\mathbf{g}(\Omega_{t})+\mathbf{J}_{t}\mathbf{M}\widetilde{\mathbf{V}}_{t}-\mathbf{M}\widetilde{\mathbf{V}}_{t}-\mathbf{T}_{0}(\mathbf{M}\widetilde{\mathbf{V}}_{t})\mathbf{e}],
\end{displaymath}
\noindent where the matrix $ \mathbf{J}_{t} $ is the matrix form of queue state transition probability $\mathrm{Pr.}[\widetilde{\mathbf{Q}}'|\mathbf{H}_{t(\mathbf{Q},\mathbf{B})},\mathbf{Q},\Omega(\mathbf{H}_{t(\mathbf{Q},\mathbf{B})},\mathbf{Q},\mathbf{B})]$ $\mathrm{Pr.}[\widetilde{\mathbf{B}}'|\mathbf{B},\Omega(\mathbf{H}_{t(\mathbf{Q},\mathbf{B})},\mathbf{Q},\mathbf{B})]$ with given $\mathbf{H}_{t(\mathbf{Q},\mathbf{B})}$ in each row, which is the real-time observation of channel state at time slot $t(\mathbf{Q},\mathbf{B})$ with state $(\mathbf{Q}_{t(\mathbf{Q},\mathbf{B})},\mathbf{B}_{t(\mathbf{Q},\mathbf{B})})=(\mathbf{Q},\mathbf{B})$. Define
\begin{displaymath}
\mathbf{Y}_{t}=\mathbf{M}^{\dag}[\mathbf{g}(\Omega_{t})+\mathbf{J}_{t}\mathbf{M}\widetilde{\mathbf{V}}_{t}-\mathbf{M}\widetilde{\mathbf{V}}_{t}-\mathbf{T}_{0}(\mathbf{M}\widetilde{\mathbf{V}}_{t})\mathbf{e}]
\end{displaymath}
\noindent and $ \delta\mathbf{Z}_{t}=\mathbf{q}_{t}-\mathbf{Y}_{t} $ and $ \mathbf{Z}_{t}=\sum_{i=j}^{t}\varepsilon_{i}^{v}\delta\mathbf{Z}_{i} $. The online value function estimation can be rewritten as
   \begin{equation}
   \label{eq37}
      \widetilde{\mathbf{V}}_{t+1}=\widetilde{\mathbf{V}}_{t}+\varepsilon_{i}^{v}\mathbf{Y}_{t}=\widetilde{\mathbf{V}}_{t}+\varepsilon_{i}^{v}\mathbf{q}_{t}-\varepsilon_{i}^{v}\delta\mathbf{Z}_{t}=\widetilde{\mathbf{V}}_{j}+\sum_{i=j}^{t}\varepsilon_{i}^{v}\mathbf{q}_{i}-\mathbf{Z}_{t}.
   \end{equation}
\noindent Our proof of Lemma 6 can be divided into the following steps:
\begin{enumerate}
  \item Step 1: Letting $ \mathcal{F}_{t}=\sigma(\mathbf{V}_{m},m \leq t) $, it is easy to see that $ \mathbf{E}[\delta\mathbf{Z}_{t}|\mathcal{F}_{t-1}]=0 $. Thus, $ \{\delta\mathbf{Z}_{t}|\forall t\} $ is a Martingale difference sequence and $ \{\mathbf{Z}_{t}|\forall t\} $ is a Martingale sequence. Moreover, $ \mathbf{Y}_{t} $ is an unbiased estimation of $ \mathbf{q}_{t} $ and the estimation noise is uncorrelated.

  \item Step 2: According to the uncorrelated estimation error from step 1, we have
  \begin{displaymath}
  \begin{split}
  &\mathbf{E}[|\mathbf{Z}_{t}|^2\big|\mathcal{F}_{j-1}]=\mathbf{E}[|\sum_{i=j}^{t}\varepsilon_{i}^{v}\delta\mathbf{Z}_{i}|^2\big|\mathcal{F}_{j-1}]=\sum_{i=j}^{t}\mathbf{E}[|\varepsilon_{i}^{v}\delta\mathbf{Z}_{i}|^2\\
  & \big|\mathcal{F}_{j-1}]=\widetilde{\mathbf{Z}}\sum_{i=j}^{t}(\varepsilon_{i}^{v})^2\rightarrow 0 \textrm{ when } j\rightarrow \infty
  \end{split}
  \end{displaymath}
  where $ \widetilde{\mathbf{Z}}\geq \max_{j\leq i\leq t}\mathbf{E}[|\delta\mathbf{Z}_{i}|^2\big|\mathcal{F}_{j-1}] $ is a bounded constant vector and the convergence of $ \widetilde{\mathbf{Z}}\sum_{i=j}^{t}(\varepsilon_{i}^{v})^2 $ is from the definition of sequence $ \{\varepsilon_{i}^{v}\} $.

  \item Step 3: From step 1, $ \{\delta\mathbf{Z}_{t}|\forall t\} $ is a Martingale sequence. Hence, according to the inequality of Martingale sequence, we have
  \begin{displaymath}
  \Pr[\sup_{j\leq i\leq t}|\mathbf{Z}_{i}|\geq \lambda \big|\mathcal{F}_{j-1}]\leq \frac{\mathbf{E}[|\mathbf{Z}_{t}|^2\big|\mathcal{F}_{j-1}]}{\lambda^2},\forall \lambda>0.
  \end{displaymath}
  From the conclusion of step 2, we have
    \begin{displaymath}
  \lim_{j\rightarrow \infty}\Pr[\sup_{j\leq i\leq t}|\mathbf{Z}_{i}|\geq \lambda \big|\mathcal{F}_{j-1}]=0,\forall \lambda>0.
  \end{displaymath}
  Hence, from \eqref{eq37}, we almost surely have $ \widetilde{\mathbf{V}}_{t+1}=\widetilde{\mathbf{V}}_{j}+\sum_{i=j}^{t}\varepsilon_{i}^{v}\mathbf{q}_{i} $ when $ j\rightarrow \infty $.
\end{enumerate}
\end{IEEEproof}

Moreover, the following lemma is about the limit of sequence $ \{\mathbf{q}_{t}\}$.

\newtheorem{lemma7}[lemma]{Lemma}
\begin{lemma7}Suppose the following two inequalities are true for $t=m,m+1,...,m+n$:
   \begin{equation}
   \label{eq38}
      \mathbf{g}(\Omega_{t})+\mathbf{P}(\Omega_{t})\mathbf{M}\widetilde{\mathbf{V}}_{t}\leq\mathbf{g}(\Omega_{t-1})+\mathbf{P}(\Omega_{t-1})\mathbf{M}\widetilde{\mathbf{V}}_{t},
   \end{equation}
   \begin{equation}
   \label{eq39}
      \mathbf{g}(\Omega_{t-1})+\mathbf{P}(\Omega_{t-1})\mathbf{M}\widetilde{\mathbf{V}}_{t-1}\leq\mathbf{g}(\Omega_{t})+\mathbf{P}(\Omega_{t})\mathbf{M}\widetilde{\mathbf{V}}_{t-1},
   \end{equation}
\noindent then we have
   \begin{displaymath}
   \lim_{t\rightarrow+\infty}\mathbf{q}_{t}=0.
   \end{displaymath}
\end{lemma7}

\begin{IEEEproof}From \eqref{eq38} and \eqref{eq39}, we have
   \begin{displaymath}
   \begin{split}
\mathbf{q}_{t}&=\mathbf{M}^{\dag}[\mathbf{g}(\Omega_{t})+\mathbf{P}(\Omega_{t})\mathbf{M}\widetilde{\mathbf{V}}_{t}-\mathbf{M}\widetilde{\mathbf{V}}_{t}-\omega_{t}\mathbf{e}]\\
  &\leq\mathbf{M}^{\dag}[\mathbf{g}(\Omega_{t-1})+\mathbf{P}(\Omega_{t-1})\mathbf{M}\widetilde{\mathbf{V}}_{t}-\mathbf{M}\widetilde{\mathbf{V}}_{t}-\omega_{t}\mathbf{e}]
   \end{split}
   \end{displaymath}
   \begin{displaymath}
   \begin{split}
   \mathbf{q}_{t-1}&=\mathbf{M}^{\dag}[\mathbf{g}(\Omega_{t-1})+\mathbf{P}(\Omega_{t-1})\mathbf{M}\widetilde{\mathbf{V}}_{t-1}-\mathbf{M}\widetilde{\mathbf{V}}_{t-1}-\omega_{t-1}\mathbf{e}] \\
   &\leq \mathbf{M}^{\dag}[\mathbf{g}(\Omega_{t})+\mathbf{P}(\Omega_{t})\mathbf{M}\widetilde{\mathbf{V}}_{t-1}-\mathbf{M}\widetilde{\mathbf{V}}_{t-1}-\omega_{t-1}\mathbf{e}]
   \end{split}
   \end{displaymath}
\noindent where $ \omega_{t}=\mathbf{T}_{0}(\mathbf{M}\widetilde{\mathbf{V}}_{t}) $. According to Lemma 6, we have
   \begin{displaymath}
   \widetilde{\mathbf{V}}_{t}=\widetilde{\mathbf{V}}_{t-1}+\varepsilon_{t-1}^{v}\mathbf{q}_{t-1},
   \end{displaymath}
\noindent therefore
   \begin{displaymath}
   \begin{split}
   \mathbf{q}_{t}&\geq[(1-\varepsilon_{t-1}^{v})\mathbf{I}+\mathbf{M}^{\dag}\mathbf{P}(\Omega_{t})\mathbf{M}\varepsilon_{t-1}^{v}]\mathbf{q}_{t-1}+\omega_{t-1}\mathbf{e}-\omega_{t}\mathbf{e} \\
   &=\mathbf{A}_{t-1}\mathbf{q}_{t-1}+\omega_{t-1}\mathbf{e}-\omega_{t}\mathbf{e}
   \end{split}
   \end{displaymath}
   \begin{displaymath}
   \begin{split}
\mathbf{q}_{t}&\leq[(1-\varepsilon_{t-1}^{v})\mathbf{I}+\mathbf{M}^{\dag}\mathbf{P}(\Omega_{t-1})\mathbf{M}\varepsilon_{t-1}^{v}]\mathbf{q}_{t-1}+\omega_{t-1}\mathbf{e}-\omega_{t}\mathbf{e}\\
   &=\mathbf{B}_{t-1}\mathbf{q}_{t-1}+\omega_{t-1}\mathbf{e}-\omega_{t}\mathbf{e}.
   \end{split}
   \end{displaymath}

\noindent Thus, we have
   \begin{displaymath}
   \mathbf{A}_{t-1}\cdot\cdot\cdot\mathbf{A}_{t-\beta}\mathbf{q}_{t-\beta}-C_1\mathbf{e}\leq\mathbf{q}_{t}\leq\mathbf{B}_{t-1}\cdot\cdot\cdot\mathbf{B}_{t-\beta}\mathbf{q}_{t-\beta}-C_1\mathbf{e}
   \end{displaymath}
   \begin{displaymath}
   \Rightarrow(1-\delta_{\beta})(\min\mathbf{q}_{t-\beta})\mathbf{e}\leq\mathbf{q}_{t}+C_1\mathbf{e}\leq(1-\delta_{\beta})(\max\mathbf{q}_{t-\beta})\mathbf{e}
   \end{displaymath}
   \begin{displaymath}
   \Rightarrow \left\{
                \begin{array}{c}
                \min\mathbf{q}_{t}+C_1\geq(1-\delta_{\beta})\min\mathbf{q}_{t-\beta}  \\
                \max\mathbf{q}_{t}+C_1\leq(1-\delta_{\beta})\max\mathbf{q}_{t-\beta}\end{array}\right.
   \end{displaymath}
   \begin{displaymath}
   \Rightarrow\max\mathbf{q}_{t}-\min\mathbf{q}_{t}\leq(1-\delta_{\beta})(\max\mathbf{q}_{t-\beta}-\min\mathbf{q}_{t-\beta})
   \end{displaymath}
   \begin{displaymath}
   \Rightarrow |\mathbf{q}_{t}^{k}|\leq \max\mathbf{q}_{t}-\min\mathbf{q}_{t}\leq C_2(1-\delta_{\beta}),\forall k
   \end{displaymath}
\noindent Then we have
   \begin{equation}
   \label{eq40}
   0\leq|\mathbf{q}_{m+n}^{k}|\leq C_3\prod_{i=0}^{\lfloor n/\beta\rfloor-1}(1-\delta_{m+i\beta})=0,\forall k.
   \end{equation}
\noindent where the first step is due to conditions on matrix sequence $ \mathbf{A}_{t} $ and $ \mathbf{B}_{t} $, $ \min\mathbf{q}_{t} $ and $ \max\mathbf{q}_{t} $ denote the minimum and maximum elements in $ \mathbf{q}_{t} $, respectively, $ \mathbf{q}_{t}^{k} $ denotes the $k$th element of the vector $ \mathbf{q}_{t} $, $|\mathbf{q}_{t}^{k}|\leq \max\mathbf{q}_{t}-\min\mathbf{q}_{t}$ is due to $\min\mathbf{q}_{t}\leq 0$, and $ C_1 $,$ C_2 $ and $ C_3 $ are constants.
\noindent According to the property of sequence $ \{\varepsilon_{t}^{v}\} $, we have
   \begin{displaymath}
   \lim_{t\rightarrow+\infty}\prod_{i=0}^{\lfloor t/\beta\rfloor-1}(1-\varepsilon_{i\beta})=0.
   \end{displaymath}
\noindent And note that $ \delta_{t}=\mathcal{O}(\varepsilon_{t}^{v}) $, from \eqref{eq37}, we have
   \begin{displaymath}
   \lim_{t\rightarrow+\infty}\mathbf{q}_{t}^{k}=0,\forall k
   \end{displaymath}
\noindent Summarize the conclusions above, we have
   \begin{displaymath}
   \lim_{t\rightarrow+\infty}\mathbf{q}_{t}=0.
   \end{displaymath}
\end{IEEEproof}
\noindent Therefore, \eqref{eq36} is straightforward when $ \mathbf{q}_{t}\rightarrow 0 $. This completes the proof.

\subsection{Proof of Lemma 4}
Due to the separation of time scale, the primal update of the per-node value function converges to $\widetilde{\mathbf{V}}_{\infty}(\boldsymbol{\eta})$ with respect to current LM $\boldsymbol{\eta}$ \cite{SystControl:Borkar}. By Lemma 4.2 in \cite{SystControl:constrained}, $G(\boldsymbol{\eta})$ is a concave and continuously differentiable except at finitely many points where both right and left derivatives exist. Since scheduling policy is discrete, we have $\Omega^{*}(\boldsymbol{\eta})=\Omega^{*}(\boldsymbol{\eta}+\bigtriangleup_{\eta})$. Thus, by chain rule, we have
   \begin{displaymath}
   \frac{\partial G(\boldsymbol{\eta}_{t})}{\partial \eta_{n,t}}=\sum _{n}\frac{\partial G(\boldsymbol{\eta}_{t})}{\partial p_{n}^{\mathrm{T}*}}\frac{\partial p_{n}^{\mathrm{T}*}}{\partial \eta_{n,t}}+\mathbf{E}^{\Omega^{*}(\boldsymbol{\eta}_{t})}[Q_{n}-D_{\mathrm{max}}r(H_{n,t},p_{n}^{\mathrm{T}})].
   \end{displaymath}
\noindent Since $\Omega^{*}(\boldsymbol{\eta}_{t})=\arg \min_{\Omega} G(\boldsymbol{\eta}_{t})$, we have
   \begin{displaymath}
   \frac{\partial G(\boldsymbol{\eta}_{t})}{\partial \eta_{n,t}}=0+\mathbf{E}^{\Omega^{*}(\boldsymbol{\eta}_{t})}[Q_{n}-D_{\mathrm{max}}r(H_{n,t},p_{n}^{\mathrm{T}})].
   \end{displaymath}
\noindent Using standard stochastic approximation theorem \cite{Cambridge:Borkar}, the dynamics of the LM update equation in \eqref{eq35} can be represented by the following ordinary differential equation (ODE):
   \begin{displaymath}
   \boldsymbol{\eta}_{t}^{'}=\mathbf{E}^{\Omega^{*}(\boldsymbol{\eta}_{t})}[Q_{n}-D_{\mathrm{max}}r(H_{n,t},p_{n}^{\mathrm{T}})].
   \end{displaymath}
\noindent Therefore, we show that the above ODE can be expressed as $\boldsymbol{\eta}_{t}^{'}=\bigtriangledown G(\boldsymbol{\eta}_{t})$. As a result, the above ODE will converge to $\bigtriangledown G(\boldsymbol{\eta}_{t})=0$, which corresponds to \eqref{eq35}. This completes the proof.

\subsection{Proof of Theorem 1}
The lower-bound is straightforward. The proof of upper-bound is given below. Since
\begin{align*}
\parallel\widetilde{\mathbf{V}}^{\infty}-\mathbf{X}^{*}\parallel&\leq\parallel\widetilde{\mathbf{T}}^{(n)}(\widetilde{\mathbf{V}}^{\infty})-\widetilde{\mathbf{T}}^{(n)}(\mathbf{X}^{*})\parallel+\parallel\widetilde{\mathbf{T}}^{(n)}(\mathbf{X}^{*})-\mathbf{X}^{*}\parallel \\
&\leq\beta\parallel\widetilde{\mathbf{V}}^{\infty}-\mathbf{X}^{*}\parallel+\parallel\widetilde{\mathbf{T}}^{(n)}(\mathbf{X}^{*})-\mathbf{X}^{*}\parallel,
\end{align*}
\noindent we have
\begin{equation}
\label{eq41}
\parallel\widetilde{\mathbf{V}}^{\infty}-\mathbf{X}^{*}\parallel\leq\frac{1}{1-\beta}\parallel\widetilde{\mathbf{T}}^{(n)}(\mathbf{X}^{*})-\mathbf{X}^{*}\parallel.
\end{equation}
\noindent From the definition of constant $c$, we have
\begin{align}
\parallel\widetilde{\mathbf{T}}^{(n)}(\mathbf{X}^{*})-\mathbf{X}^{*}\parallel&\leq\parallel\widetilde{\mathbf{T}}^{(n)}(\mathbf{X}^{*})-\mathbf{M}^{\dagger}\mathbf{V}^{\infty}\parallel+\parallel\mathbf{M}^{\dagger}\mathbf{V}^{\infty}-\mathbf{X}^{*}\parallel \nonumber \\
&\leq c\parallel\widetilde{\mathbf{T}}^{(n-1)}(\mathbf{X}^{*})-\mathbf{X}^{*}\parallel+\parallel\mathbf{M}^{\dagger}\mathbf{V}^{\infty}-\mathbf{X}^{*}\parallel \nonumber \\
&\leq (c^{n}+1)\parallel\mathbf{M}^{\dagger}\mathbf{V}^{\infty}-\mathbf{X}^{*}\parallel. \label{eq42}
\end{align}
\noindent As a result,
\begin{align*}
\parallel\mathbf{M}\widetilde{\mathbf{V}}^{\infty}-\mathbf{V}^{\infty}\parallel&\leq\parallel\mathbf{M}\widetilde{\mathbf{V}}^{\infty}-\mathbf{M}\mathbf{X}^{*}\parallel+\parallel\mathbf{M}\mathbf{X}^{*}-\mathbf{V}^{\infty}\parallel \\
&\leq a\parallel\widetilde{\mathbf{V}}^{\infty}-\mathbf{X}^{*}\parallel+\parallel\mathbf{M}\mathbf{X}^{*}-\mathbf{V}^{\infty}\parallel \\
&\leq\frac{a(c^{n}+1)}{1-\beta}\parallel\mathbf{M}^{\dagger}\mathbf{V}^{\infty}-\mathbf{X}^{*}\parallel+\parallel\mathbf{M}\mathbf{X}^{*}-\mathbf{V}^{\infty}\parallel,
\end{align*}
\noindent where the last inequality is because of \eqref{eq41} and \eqref{eq42}. This completes the proof.


\end{document}